\documentclass[journal]{IEEEtran}
\usepackage{amsmath,amsfonts}
\usepackage{algorithmic}
\usepackage{algorithm}
\usepackage{array}
\usepackage[caption=false,font=normalsize,labelfont=sf,textfont=sf]{subfig}
\usepackage{textcomp}
\usepackage{stfloats}
\usepackage{url}
\usepackage{verbatim}
\usepackage{graphicx}
\usepackage{cite}
\usepackage{float}
\usepackage{stfloats}
\usepackage{tabularx}
\usepackage{adjustbox}
\usepackage{lscape}
\usepackage{fullpage}
\usepackage{xtab,booktabs}

\usepackage{floatpag}
\usepackage[center]{caption}
\usepackage{tikz}
\usepackage[affil-it]{authblk} 
\usepackage{etoolbox}
\usepackage{lmodern}
\usepackage{xtab}
\usetikzlibrary{mindmap,trees}

\usepackage{fancyhdr}

\fancypagestyle{first}{%
\fancyhf{} 

\lhead{IEEE Transactions on Network and Service Management} 
\setlength{\headsep}{0.5in}

}
\begin{document}

\title{A Survey on Explainable Artificial Intelligence for  Cybersecurity
}

\author{
\IEEEauthorblockN{Gaith Rjoub\IEEEauthorrefmark{1}\IEEEauthorrefmark{2}, Jamal Bentahar\IEEEauthorrefmark{2}\IEEEauthorrefmark{4}, Omar Abdel 
Wahab\IEEEauthorrefmark{3}, 
Rabeb Mizouni\IEEEauthorrefmark{5}, 
Alyssa Song\IEEEauthorrefmark{6}, 
Robin Cohen\IEEEauthorrefmark{6}, 
Hadi Otrok\IEEEauthorrefmark{5}, and Azzam Mourad\IEEEauthorrefmark {7} \IEEEauthorrefmark{8}}\\
    
\IEEEauthorblockA{\IEEEauthorrefmark{1} \normalsize King Hussein School of Computing Sciences, Princess Sumaya University for Technology, Jordan}\newline
\IEEEauthorblockA{\IEEEauthorrefmark{2} Concordia Institute for Information Systems Engineering, Concordia University, Montreal, Canada}\newline
\IEEEauthorblockA{\IEEEauthorrefmark{3}Department of Computer and Software Engineering, Polytechnique Montréal, Montréal, Canada}\newline
\IEEEauthorblockA{\IEEEauthorrefmark{4}Electrical Engineering and Computer Science Department, Khalifa University, Abu Dhabi, UAE}\newline
\IEEEauthorblockA{\IEEEauthorrefmark{5}Center of Cyber-Physical Systems (C2PS), Department of EECS, Khalifa University, Abu Dhabi, UAE}\newline
\IEEEauthorblockA{\IEEEauthorrefmark{6}David R. Cheriton School of Computer Science, University of Waterloo, Canada}\newline 
\IEEEauthorblockA{\IEEEauthorrefmark{7}Cyber Security Systems and Applied AI Research Center, Dept. of CSM, Lebanese American University}\newline
\IEEEauthorblockA{\IEEEauthorrefmark{8}  Division of Science, New York University, Abu Dhabi, UAE}\newline

 Email:   g.rjoub@psut.edu.jo, bentahar@ciise.concordia.ca, omar.abdul-wahab@polymtl.ca, rabeb.mizouni@ku.ac.ae, a8song@uwaterloo.ca, 
    rcohen@uwaterloo.ca,
    hadi.otrok@ku.ac.ae, azzam.mourad@lau.edu.lb}

\maketitle
\thispagestyle{first}

\begin{abstract}
The “black-box” nature of artificial intelligence (AI) models has been the source of many concerns in their use for critical applications. Explainable Artificial Intelligence (XAI) is a rapidly growing research field that aims to create machine learning models that can provide clear and interpretable explanations for their decisions and actions. In the field of cybersecurity, XAI has the potential to revolutionize the way we approach network and system security by enabling us to better understand the behavior of cyber threats and to design more effective defenses. In this survey, we review the state of the art in XAI for cybersecurity  and explore the various approaches that have been proposed to address this important problem. The review follows a systematic classification of cybersecurity threats and issues in networks and digital systems. We discuss the challenges and limitations of current XAI methods in the context of cybersecurity and outline promising directions for future research.
\end{abstract}


\section{Introduction}
\IEEEPARstart{A}{rtificial}
Intelligence (AI), and in particular its rapidly growing sub-field of deep learning, has become a dominant force in both research and industrial applications in the domain of computer science and beyond. This is particularly true given the undeniable benefits that AI has brought to many critical domains such as digital health, transportation and manufacturing. 
However, many AI methods and algorithms have become increasingly difficult to understand and explain, making it difficult to answer many crucial questions such as (1) who is responsible if things go wrong$?$ (2) can we explain the output and internal structure of the AI method$?$ (3) Can we trust the decisions generated by the underlying AI method and why$?$.

Motivated by these questions, the concept of eXplainable Artificial Intelligence (XAI) has arisen in the past few years in an attempt to address the explainability issues of Machine Learning (ML) algorithms. 
By combining new or modified ML techniques with effective explanation techniques, XAI aims to create an explanation-enriched approach to ML that enables end users to understand, trust, and effectively manage the upcoming generation of AI systems. AI systems are usually capable of making decisions, making recommendations, and taking actions, but they need to justify their reasoning to end users. Analysts, for instance, must understand the intuition behind a recommendation from a big data analytics system for further investigation. Similarly, an operator who intends to use some autonomous vehicles should be aware of how they make their driving decisions. XAI tries to provide answers for these questions by proposing a set of tools and concepts that can be integrated into existing ML techniques. In summary, XAI mainly allows us to:

\begin{itemize}
    \item Rationally justify the model's decisions and detect its errors to enhance its reliability and safety,
    especially during the trial period of an AI model.
    \item Adhere to regulations to provide a transparency that enables accountability, improved security, and the protection of personal information.
    \item Enhance our understanding of AI reasoning and mitigate the fairness concerns associated with its use.
    \item Help practitioners validate AI systems' proprieties to ensure they meet the developer's requirements.
\end{itemize}

\subsection*{Motivation and Contributions}
In the domain of cybersecurity, the application of AI has become mainstream in both research and industry. This interest is largely driven by the fact that AI-based tools and methods are claimed to be able to detect and prevent sophisticated attacks in networks and digital systems that could not be detected and prevented by conventional approaches. However, such tools and methods also raise concerns regarding their potential misuse, especially in critical domains such as cybersecurity. This opens up a debate on the need for AI tools that are able to explain their behavior. With the aim of providing a better understanding of the importance and need for XAI in the domain of cybersecurity, we survey in this paper state-of-the-art XAI, with a particular focus on XAI approaches applicable to cybersecurity. We provide an overview on the general principles and methodologies for XAI literature and analyze their potential benefits when applied to cybersecurity. We then provide a comprehensive classification of the cybersecurity domain, which we use to discuss the literature and promising applications of XAI in each cybersecurity sub-domain. Finally, we discuss promising directions for future research and some ethical challenges associated with XAI for cybersecurity.

The future of cybersecurity will be defined by how we approach AI models and how we can make them safer, transparent and efficient through the use of XAI solutions. For example, ML models that can explain their decisions can reduce the number of false alarms, which can lead to better security for users. They can also reduce the need for human oversight, which can save money for organizations. As we move toward a world where we can program our machines to make decisions for us, we will need to build systems that can explain their actions to humans when appropriate. It is essential, despite the fact that this is one of the very few surveys on XAI for cybersecurity, 
that we begin this discussion with the caveat that there is a long way to go before we see widespread use of XAI in the real world for the purpose of ensuring network and system security.  To analyze the impact of this fact on the field and discuss how it can be addressed in the future, it is necessary to identify the various ways in which XAI can be used to improve cybersecurity. This can be done by examining existing products and services that are based on XAI, as well as the proposed use cases. It is also important to discuss the challenges that have been faced in the field of  cybersecurity, and how XAI can be used to address them. More specific contributions of this article are reported in Section \ref{related} when we compare our survey with the related ones.

The remainder of the paper is organized as follows. Preliminary concepts important for the rest of the paper are explained in Section \ref{Preliminaries}. A detailed description of the systematic methodology used to search for relevant research articles is provided in Section \ref{Methodology}.  Section \ref{related} presents related review papers and discusses similarities and differences with the current survey paper, which helps us identify the unique characteristics and the value added of our survey. A Classification of the selected scientific articles on explainability of AI models for cybersecurity based on the proposed solutions is provided in Section \ref{Papers_Classification}.
Section \ref{Math_Models} discusses the use of mathematical models for explainability in AI models for cybersecurity.
Section \ref{xai-methods} provides a classification of XAI methods and techniques for cybersecurity.
Section \ref{xai-cyb} explores the use of XAI-based solutions for  cybersecurity.
Section \ref{criteria} introduces a set of desirable criteria for future solutions for explainability in AI models for cybersecurity. Section \ref{future} outlines potential directions for future work in the field of explainability in AI models for cybersecurity.
Finally, Section \ref{conc} concludes the paper and summarizes the key findings and contributions of the study.

\section{Preliminaries} \label{Preliminaries}
Before delving into the literature of XAI-based cybersecurity, let us describe key terms that will be used throughout this article. It is important to distinguish between the terms of explainability, interpretability, trustworthiness, interactivity, stability, and robustness as they are very related and often used interchangeably when in fact, they have different meanings. 

\begin{itemize}
    \item \textbf{Explainability}: Explainability refers to the degree to which an AI system is able to explain its decisions and act on them despite being opaque to an observer \cite{gevaert2022explainable}. Explainability is especially important for AI systems that are used for task automation \cite{kumara2022focloud}. Without explainability, it is impossible to know whether or not the system is making decisions in a way that is “fair".  
    \item \textbf{Interpretability}: Interpretability is the degree to which an AI system is able to provide insights into its decision-making process \cite{erasmus2021interpretability,carletti2023interpretable}. It can be measured by the amount of information that the system is able to provide to help users understand how the system is making its decisions. Interpretability is also important for ensuring that people can trust the decisions of the AI solution.
    \item \textbf{Trustworthiness}: Trustworthiness refers to the degree to which a person, an organization, or a system is perceived as honest, reliable, and ethical \cite{polley2021towards,yan2020trustworthy,rjoub2020bigtrustscheduling,wahab2020endorsement,drawel2021formalizing,wahab2022federated}. It is important to note that trustworthiness can be viewed as a consequence of both interpretability and explainability. It is also crucial to note that trustworthiness is not synonymous with honesty and ethics, whereas honesty and ethics are merely the outward manifestations of trustworthiness \cite{AlwhishiBE22,BentaharDS22,DrawelLBE22,MousaBA21}.
    \item \textbf{Interactivity}: Interactivity refers to the degree to which a system is able to interact with its users \cite{sokol2020one,cui2020hybrid}. It has been argued that making AI systems more interactive helps  increase both trustworthiness and interpretability.
    \item \textbf{Stability}: 
    Stability in AI refers to the degree to which an AI system can maintain its intended functionality under a variety of conditions. The goal of designing stable AI systems is to ensure that the system is robust, reliable, and consistent in its performance, even in the presence of unexpected inputs or changes in the operating environment. This is particularly important for AI systems that are used in critical applications, such as autonomous vehicles or medical diagnosis systems, as it helps designers minimize the risk of system failure or unintended behavior. The stability of an AI system depends on many factors, including the quality of the training data, the design of the algorithms and models, and the underlying hardware and software infrastructure. By prioritizing stability in AI system design, organizations can help in building trustworthy, reliable, and safe AI systems that can deliver consistent and valuable outcomes.
   \item \textbf{Robustness}: Robustness refers to the ability of an AI system to continue to perform well despite being subjected to changes in input \cite{ali2022tamp,li2020read}. It is often used to describe the stability of AI systems that are used by robots and other autonomous systems to perform tasks in the real world \cite{valluripally2022detection}. 
\end{itemize}

\section{Research Methodology} \label{Methodology}
In recent years, XAI has been increasingly used in cybersecurity, with researchers and practitioners exploring the potential for AI to detect and explain cybersecurity errors. In many ways, the field of XAI is reminiscent of the literature on human error, with researchers and practitioners exploring the same sorts of questions around human error—how it arises, i.e., how to prevent it, and how to detect it. However, there is one key difference between the literature on human error and the literature on XAI: human errors may be deliberate and often malicious, while cybersecurity errors can be accidental. The field of network and system security primarily focuses on prevention of cybersecurity errors rather than detection and explanation. This is because it is difficult to set clear boundaries for detecting and explaining errors in a precise and indisputable way. Instead, the focus is on preventing malicious or deliberate errors through network and system security measures such as firewalls, encryption, and intrusion detection systems.
This is due to the multidisciplinary nature of this fascinating new field of research. It spans from computer science to mathematics, psychology to human factors, and philosophy to Ethics. Computer science, statistics, and mathematics are associated primarily with the development of computational models from data, whereas human factors and psychology are more related to the study of explainability since human beings are involved. Ethics and philosophy are intertwined when discussing explainability.
Due to this, for our survey, we imposed some constraints, and the following types of publications were excluded from consideration:

\begin{itemize}
    \item Studies relating to the concept of explainability in other disciplines than artificial intelligence or cybersecurity;
    \item Non-peer-reviewed articles or technical reports;
    \item Studies designed only to improve model transparency and not focused specifically on explanation.
\end{itemize}
This systematic review was conducted in two phases, taking into account the constraints discussed above:

\begin{itemize}
\item The first phase was to identify the major categories of studies on explainability in artificial intelligence and cybersecurity. This was done by systematically searching the literature databases (i.e. Scopus, IEEE, Springer, Sience Direct, Google Scholar) where the search was limited to peer-reviewed journal articles, book chapters, and conference proceedings.
\item In the second phase, we scanned the reference lists of all relevant publications in order to identify papers that met the inclusion criteria. We excluded studies that were not about explainability in cybersecurity and AI.
\end{itemize}

\begin{figure}[!htp]
\includegraphics[width=0.5\textwidth,height=0.32\textwidth]{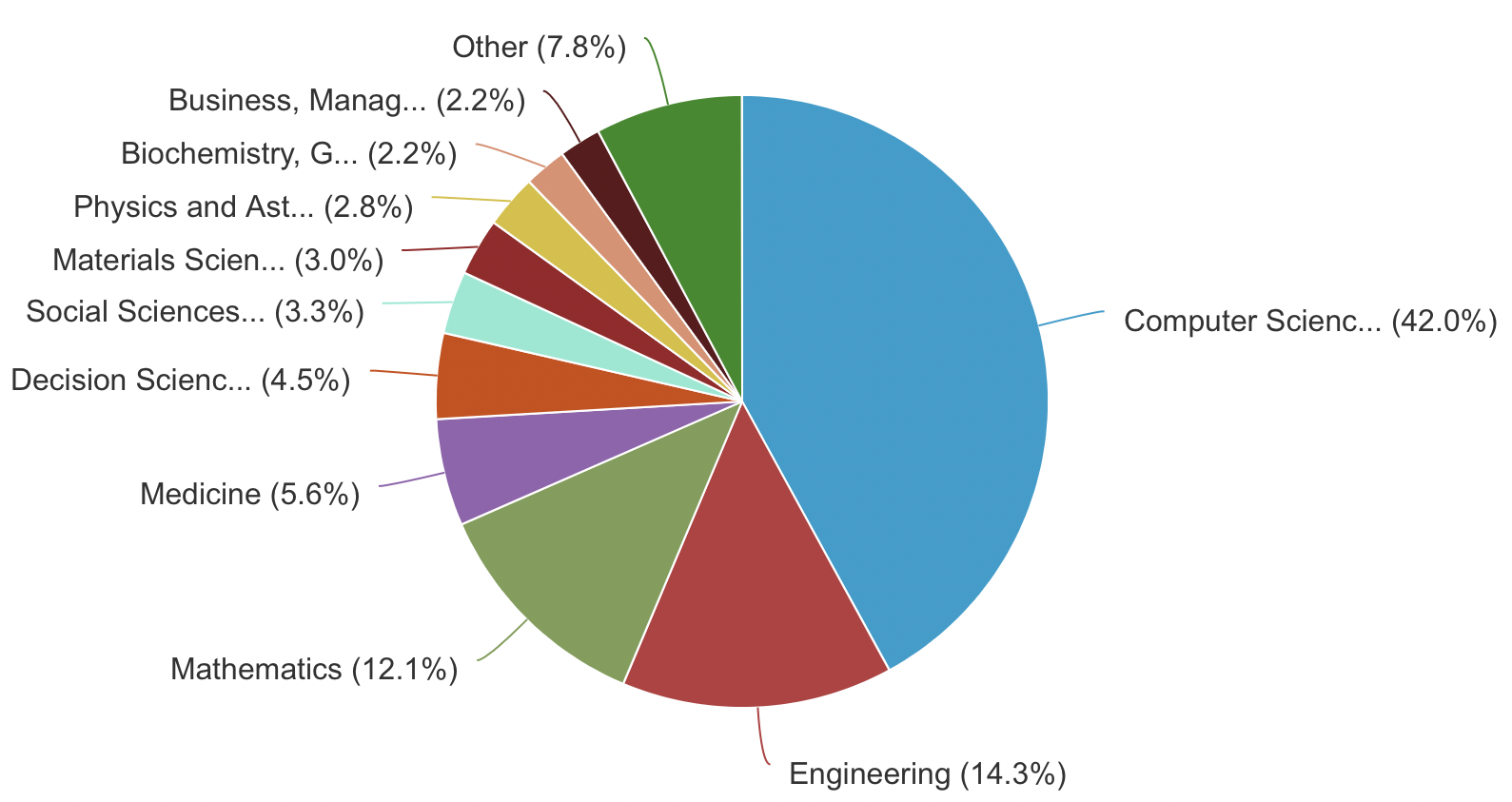}
\caption{Percentage breakdown of the XAI literature based on the publication subject area}
\label{fig1}
\end{figure}

\begin{figure}
 \includegraphics[width=.5\textwidth,height=.32\textwidth]{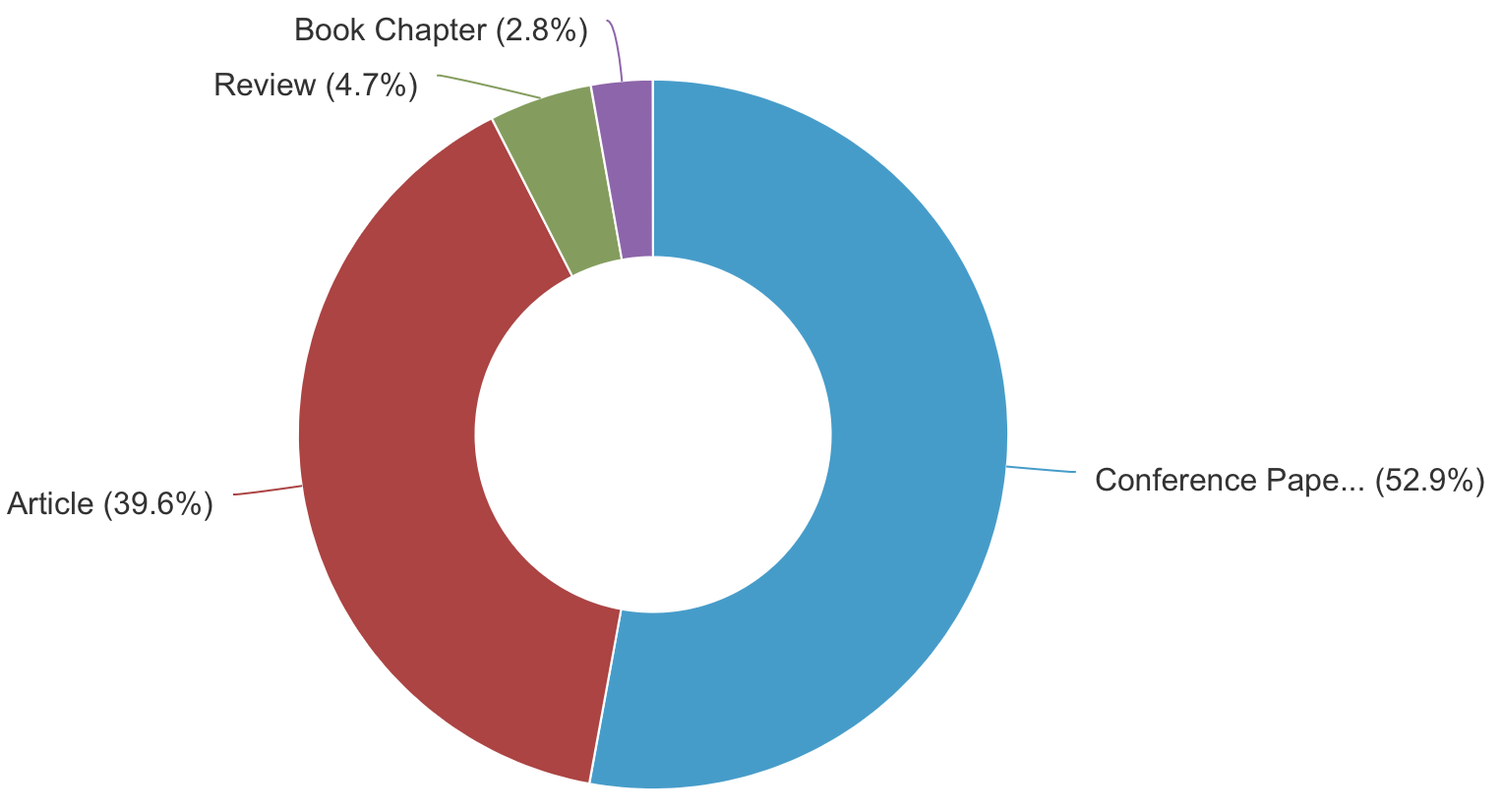}
  \caption{Percentage breakdown of the XAI literature based on the publication type}
  \label{fig2}
\end{figure}
According to the type and subject areas of the publications, Fig. \ref{fig1} and Fig. \ref{fig2} show the percentage breakdown of the XAI-based cybersecurity literature. In fact, $42\%$ of all XAI-based security articles are published as conference or journal articles ($52.9\%$ and $39.6\%$ respectively). On the other hand, review papers account for $4.7\%$ of the papers while book chapters account for $2.8\%$.

\section{Literature Review}
\label{related}
Recently, a few survey articles have been published on  XAI. In this section, we review these surveys and highlight the unique contributions of our work. In \cite{alicioglu2022survey}, the authors present a survey that explores the current trends and challenges of using visual analytics to interpret deep learning models based on XAI methods, as well as future directions of research in this area. Two perspectives have been taken into consideration, model usage and visual approaches. Model usage focuses on the performance of the AI system and how it behaves in different scenarios, while visual approaches examine the behavior of the AI system by looking at its outputs, such as predictions or classifications, over time. The researchers identified several research questions in light of their findings, then discussed the research directions that could be pursued in the future. By using XAI methods in the field of visual analytics, this survey provides guidance to better interpret neural networks.
Researchers in \cite{tjoa2020survey} provide a survey of interpretability and explainability related to ML algorithms, categorizing different interpretations proposed by different research works. Their categorization also pertains to medical applications. According to the researchers, interpretability research can broadly be classified in many ways. It can range from methods that provide clearly interpreted information to analysis of complex patterns. The authors present attempts to mathematically formalize interpretability, as well as efforts to visualize it and evaluate its impact on task performance. These different approaches aim to better understand and improve the interpretability of algorithms.


The investigators in \cite{srivastava2022xai} provide a literature review of the use of XAI in the field of cybersecurity. The authors first discuss the current state of research in XAI and provide an overview of the different approaches and techniques that have been used to develop explainable models. They then explore the potential applications of XAI in cybersecurity, including the detection of cyberattacks, the identification of vulnerabilities, and the improvement of decision-making in security operations.
The paper also identifies several challenges and open issues in the application of XAI to cybersecurity, such as the need for effective methods for validating and testing XAI models, the challenge of balancing interpretability and accuracy in model development, and the importance of developing XAI models that can adapt to evolving threats and attack techniques. Finally, the paper suggests some future research directions in this area, such as the development of hybrid approaches that combine XAI with other techniques such as deep learning, and the exploration of new ways to visualize and explain complex XAI models to human analysts.
The authors in \cite{capuano2022explainable} present a survey of the research and applications of XAI in the field of cybersecurity. The authors first define the concept of XAI and explain its importance in the context of cybersecurity, highlighting the need for models that can be understood and trusted by human analysts.
The paper then reviews the existing research on XAI in cybersecurity, including the different approaches and techniques that have been used to develop explainable models. The authors discuss the advantages and limitations of each approach and highlight the key challenges that need to be addressed in order to develop effective XAI models for cybersecurity.
The paper also presents an overview of the different applications of XAI in cybersecurity, such as the detection of malware, the identification of vulnerabilities in systems, and the improvement of decision-making in security operations. The authors discuss the potential benefits of XAI in each of these areas and highlight some of the key research questions that need to be addressed in order to fully realize the potential of XAI in cybersecurity.

The researchers in \cite{charmet2022explainable} provide a comprehensive review of the research on XAI and its applications in cybersecurity. The paper highlights the growing need for AI models that can be understood and interpreted by humans in the context of cybersecurity. The authors cover the applications of XAI in cybersecurity, including intrusion detection and malware classification, as well as the security of XAI pipelines. The survey explores the methods and approaches that are relevant in applying explainability in cybersecurity applications. Then, it analyzes the security of XAI methods, identifies trends and challenges in the field. The survey proposes technical research avenues that would contribute to the advancement of AI and XAI in the cybersecurity domain. The paper also identifies open research questions and future directions for XAI in cybersecurity.
In \cite{zhang2022explainable}, the authors focus on the applications of XAI in the field of cybersecurity, including model-agnostic techniques, rule-based approaches, and hybrid methods. The paper highlights the limitations of conventional rule-based and signature-based cybersecurity strategies and how the use of AI, including ML and DL, can enhance the detection and defense of cyber attacks, such as intrusion detection, malware detection, and spam filtering. However, the "black-box" nature of most AI-based techniques presents a challenge in understanding how these models reach certain conclusions, reducing human users' confidence in these models' efficacy. Therefore, the paper emphasizes the need for XAI methods in establishing cybersecurity models that are transparent and interpretable to human users while maintaining high accuracy. The survey aims to review different techniques, existing challenges, frameworks, and datasets for XAI-based cyber defense mechanisms. Additionally, it reviews successful XAI-based systems and applications in the cybersecurity domain and identifies challenges and research gaps in XAI applications in cybersecurity. Finally, the paper presents key insights and future research directions for applying XAI in the cyber security area.

The researchers in \cite{das2020opportunities} present mathematical explanations of significant publications in addition to providing a complete picture of the current XAI environment in deep learning. They present a taxonomy and categorise XAI approaches based on their scope of explanations, methodology underlying the algorithms, and explanation level or usage, which aids in the development of trustworthy, interpretable, and self-explanatory deep learning models. The authors then outline the basic principles employed in XAI research and give a historical chronology for significant XAI studies from $2007$ to $2020$. The authors thoroughly describe each type of algorithms and methodologies before evaluating the explanation maps produced by eight XAI algorithms on image data, discussing the limits of each methodology, and suggesting potential future routes for XAI assessment. In \cite{danilevsky2020survey}, the investigators  provide an overview of the current state of XAI, which is regarded within the area of Natural Language Processing (NLP). The researchers address the different categories of explanations, as well as the numerous method explanations that may be arrived at and depicted. They describe the operations and explainability strategies available for creating explanations for NLP model predictions in order to serve as a resource for model developers in the community. Finally, they highlight current gaps and suggest future routes for study in this critical field. In \cite{li2020survey}, the authors offer a survey that examines and taxonomizes existing initiatives from the perspective of Data and Knowledge Engineering (DKE), summarising their contributions, technical essences, and comparative features. They divide approaches into two categories: data-driven methods, which rely on task-related data for explanation, and knowledge-aware methods, which include extraneous knowledge. Furthermore, they present a study of the state-of-the-art assessment metrics and deployed explanation applications in industrial settings. In \cite{confalonieri2021historical}, the authors give a historical overview of XAI. They address how explainability was previously imagined, how it is currently understood, and how it could be understood in the future. They wrap up the work by suggesting an explanation criteria that they feel will be critical in the creation of human-understandable explainable systems.

The authors in \cite{mohseni2021multidisciplinary} propose a survey and methodology designed to share information and experiences about XAI design and assessment approaches across many disciplines. They provide a classification of XAI design objectives and assessment techniques following a thorough analysis of XAI related articles in the disciplines of ML, visualisation, and human-computer interaction, with the purpose of supporting different design goals and evaluation approaches in XAI research. Their classification shows the relationship between design goals for various XAI user groups and assessment techniques. They create a framework with step-by-step design principles and assessment procedures to conclude the iterative design and evaluation cycles in multidisciplinary XAI teams. Furthermore, they give ready-to-use summary tables of evaluation methodologies and recommendations for various XAI proposals. As a reference for both theorists and practitioners, the authors in \cite{linardatos2020explainable} provide a literature review and taxonomy of ML interpretability methods, along with links to their programming implementations. A taxonomy of interpretability methods was developed, identifying four major categories: methods for explaining complex black-box models, methods for creating white-box models, methods that promote fairness and prevent discrimination, and methods for analyzing the sensitivity of model predictions. A brief overview of the current state-of-the-art in relation to the explainability of AI techniques is provided in \cite{angelov2021explainable} by using recent advances in ML and deep learning in particular. Using the recently formulated four principles of explainability from the National Institute of Standards, the paper begins with a brief historical introduction and taxonomy, formulating the main challenges in terms of explainability. Afterwards, recent methods related to the topic are critically reviewed and analyzed. 

 
As demonstrated in this section, XAI is a rapidly evolving field, and there are already a few surveys conducted to understand the current state of XAI in cybersecurity. While these surveys provide valuable insights into the development and use of XAI techniques in cybersecurity, they are not without limitations. One of the main limitations of existing XAI cybersecurity surveys is that they often focus on a specific XAI technique or a subset of XAI techniques, rather than providing a comprehensive overview of the entire field that cover the various XAI classes. This can lead to a narrow view of XAI and its potential applications in cybersecurity.
Another limitation is that many surveys rely on self-reporting from developers and researchers, which can lead to bias and inaccuracies in the data. Additionally, many surveys do not take into account the diversity of stakeholders involved in XAI cybersecurity, such as end-users, policymakers, and regulators.
Finally, there is a lack of standardization in the metrics and evaluation criteria used in XAI cybersecurity surveys, particularly network security, which makes it difficult to compare results across studies and to draw definitive conclusions about the effectiveness of different XAI techniques.
Despite these limitations, XAI cybersecurity surveys remain an important tool for understanding the state of the field and identifying areas for future research and development. As the field continues to evolve, it is likely that surveys will become more sophisticated, comprehensive but focused, providing more detailed insights into the development and use of XAI in particular cybersecurity fields.

The survey we present in this article aims to provide a comprehensive review of the current literature on XAI methods for  cybersecurity applications, including a review of existing challenges and problems, identification of available frameworks and datasets, and analysis of successful XAI-based systems and applications. Its ultimate goal is to identify research gaps and future directions for applying XAI in the cybersecurity domain. Compared to the other surveys discussed in this section, our survey offers several unique features and advantages. \textbf{The first aspect} that sets this survey apart from related surveys that have a broader focus on different applications of XAI is its specific focus on the application of XAI methods in the field of  cybersecurity. This specific focus on the intersection of XAI and  cybersecurity allows us to offer a more targeted and comprehensive analysis of this particular domain, which will help researchers, network and system security engineers and practitioners identify the existing solutions and gaps. \textbf{Second}, the survey introduces two unique and systematic mappings linking eight XAI model classes that we discuss in this survey to nine desirable criteria that we identify in this article for XAI cybersecurity solutions. The first mapping is linking the eight comprehensive XAI models to our taxonomy of six cybersecurity classes. The second mapping is about linking the identified six classes of  cybersecurity to the nine assessment criteria of cybersecurity XAI solutions. The two mappings allows us to introduce a more organized and structured overview of the different approaches and techniques used in the field of cybersecurity. Combining the two mappings provides a clear understanding of which XAI models are applicable to which cybersecurity classes, and how they can be used to address specific network and system security challenges and threats. This is a unique contribution compared to the existing surveys as it helps researchers and practitioners navigate the complex landscape of XAI and cybersecurity and make informed decisions about which techniques to use in specific scenarios. Moreover, identifying and mapping the different cybersecurity classes to desirable criteria of explainable AI for cybersecurity provide a clear and concise framework for evaluating and comparing different XAI models and their suitability for different network and system security applications. 
This mapping allows for a more thorough evaluation of the XAI models in terms of their ability to meet specific criteria and address the unique challenges of cybersecurity. This approach can also aid researchers and practitioners in selecting the most appropriate XAI model for a particular cybersecurity task based on the desired outcomes and criteria. \textbf{Third}, our survey provides a detailed discussion of the limitations and challenges of current XAI approaches in cybersecurity, as well as potential solutions and future research directions.
Overall, our survey offers a unique and valuable contribution to the existing literature on XAI and cybersecurity, by providing a comprehensive and structured analysis of the current state-of-the-art, as well as identifying key areas for future research and development. We summarize in Table \ref{Table-Comparative1} the main similarities and differences between our survey and the existing surveys on XAI.

\begin{table*}
\thisfloatpagestyle{empty}
\centering
\caption{COMPARATIVE SUMMARY BETWEEN OUR SURVEY AND THE EXISTING SURVEYS ON XAI}
\label{Table-Comparative1}
\begin{tabular}{c c c}
\hline
Approach & Similarities & Differences\\ 
\hline\\
Alicioglu et al. \cite{alicioglu2022survey}&
\begin{tabular}{l}
\begin{minipage}[t]{0.8\columnwidth}%
Similar to our survey, this survey addresses how to interpret deep learning models based on XAI methods.  %
\end{minipage}\tabularnewline
\end{tabular}&\begin{tabular}{l}
\begin{minipage}[t]{0.8\columnwidth}
\begin{itemize}
\item The classification scheme proposed in this survey is based on visualization techniques that help present the model and prediction explanations. 
\item Our work addresses additional challenges that are not mentioned in \cite{alicioglu2022survey} such as intrusion prevention, access control, and privacy.
\item No classification is provided for the techniques that are proposed to address the discussed challenges.
\end{itemize}
\end{minipage}\tabularnewline
\end{tabular} \\\\\\
Tjoa et al. \cite{tjoa2020survey} & \begin{tabular}{l}
\begin{minipage}[t]{0.8\columnwidth}
Similar to our survey, this survey includes some mathematical formulations of common or popular XAI methods.
\end{minipage}\tabularnewline
\end{tabular}  & \begin{tabular}{l}
\begin{minipage}[t]{0.8\columnwidth} \begin{itemize}
\item This survey is mainly dedicated to the healthcare informatics community, while our survey is dedicated to the security community.
\item The presented classification scheme is not based on a clear criterion, which might be confusing for the reader. For example, trust and transparency are challenges of XAI, while the scale and motivation of the XAI are rather system design choices. 
\item Our classification scheme is systematic, multi-level and based on clear criteria at each level.
\end{itemize}\end{minipage}\tabularnewline
\end{tabular}\\\\\\

Srivastava al. \cite{srivastava2022xai} & \begin{tabular}{l}
\begin{minipage}[t]{0.8\columnwidth}
Similar to our survey, this survey provide an overview of XAI in cybersecurity, discussing approaches and potential applications. They identify challenges such as validating models, balancing interpretability and accuracy, and adapting to evolving threats.
\end{minipage}\tabularnewline
\end{tabular}  & \begin{tabular}{l}
\begin{minipage}[t]{0.8\columnwidth} \begin{itemize}
\item This survey is not intended for any specific group but aims to encompass policymakers, ethicists, and members of the general public who wish to comprehend the implications of XAI in cybersecurity. Our primary focus is however on researchers and practitioners in the cybersecurity field who want to understand the current state of XAI models for network and system security.  
\item Our survey delves deeply into the technical details of various XAI models, including their strengths and weaknesses, their performance on different types of cybersecurity tasks, and the specific features of network data that they are able to analyze effectively. \cite{srivastava2022xai}, in contrast, places more emphasis on the broader societal and ethical issues related to XAI in cybersecurity, such as the potential impact of these technologies on privacy, fairness, and accountability.
\end{itemize}\end{minipage}\tabularnewline
\end{tabular}\\\\\\

Capuano al. \cite{capuano2022explainable} & \begin{tabular}{l}
\begin{minipage}[t]{0.8\columnwidth}
Similar to our survey, this survey provides a comprehensive list of XAI techniques and algorithms that can be used in cybersecurity applications.
\end{minipage}\tabularnewline
\end{tabular}  & \begin{tabular}{l}
\begin{minipage}[t]{0.8\columnwidth} \begin{itemize}
\item This survey classifies XAI models into four categories based on their interpretability: black box, glass box, gray box, and white box models. The authors provide an overview of various XAI models within each of these categories and discuss their strengths and limitations.

In contrast, our survey includes a wider range of classification methods and techniques used in XAI for cybersecurity models. Additionally, our survey includes a detailed discussion of the strengths and limitations of each method, which provides readers with a comprehensive understanding of the different approaches.
\item Our work includes more XAI methods in a systematic way and also covers more recent papers published in $2021$ and $2022$.
\end{itemize}\end{minipage}\tabularnewline
\end{tabular}\\\\\\
 
\end{tabular}

\end{table*}

\begin{table*}
\thisfloatpagestyle{empty}
\centering
\begin{tabular}{c c c }
\hline
Approach & Similarities & Differences\\ 
\hline\\

Charmet al. \cite{charmet2022explainable} & \begin{tabular}{l}
\begin{minipage}[t]{0.8\columnwidth}
Similar to our survey, this survey cover the categories of model-agnostic explanations and model-specific explanations, highlighting the importance of interpretability in XAI for cybersecurity.
\end{minipage}\tabularnewline
\end{tabular}  & \begin{tabular}{l}
\begin{minipage}[t]{0.8\columnwidth} \begin{itemize}
\item Our survey uses a classification approach to organize the XAI models and techniques for cybersecurity, while the work in \cite{charmet2022explainable}  uses a narrative approach to summarize the existing literature on XAI in cybersecurity in general. 
\item Our survey places a greater emphasis on recent developments and emerging trends in XAI models for cybersecurity, while the work in \cite{charmet2022explainable}  has a historical perspective on the evolution of XAI in cybersecurity.

\end{itemize}\end{minipage}\tabularnewline
\end{tabular}\\\\\\

Zhang al. \cite{zhang2022explainable} & \begin{tabular}{l}
\begin{minipage}[t]{0.8\columnwidth}
Similar to our survey, this survey cover a range of XAI techniques and applications in the field, and provide insights into the potential benefits and challenges of using XAI models in cybersecurity.
\end{minipage}\tabularnewline
\end{tabular}  & \begin{tabular}{l}
\begin{minipage}[t]{0.8\columnwidth} \begin{itemize}
\item Our survey places greater emphasis on the challenges and limitations of XAI models for cybersecurity and provides future directions for research in this area, while the work in \cite{zhang2022explainable} is more descriptive and less prescriptive.

\item Our survey aims to provide a comprehensive classification of XAI models for cybersecurity, focusing on the types of models, techniques, and evaluation metrics used. It also covers a wide range of XAI models and their applications in network and system security. In contrast, the work in \cite{zhang2022explainable} primarily focuses on summarizing and analyzing the results of individual studies and experiments that have used XAI models for cybersecurity.

\end{itemize}\end{minipage}\tabularnewline
\end{tabular}\\\\\\

Das et al. \cite{das2020opportunities}&\begin{tabular}{l}
\begin{minipage}[t]{0.7\columnwidth}
Similar to our survey, this survey provides a multi-level classification scheme for XAI challenges, including high-level challenges and sub-challenges.
\end{minipage}\tabularnewline
\end{tabular} & \begin{tabular}{l}
\begin{minipage}[t]{0.8\columnwidth} \begin{itemize}
\item This work provides a system-level classification scheme of XAI that consists of three aspects, namely: scope, methodology, and usage.
\item Our work addresses challenges of XAI that are not discussed in \cite{das2020opportunities} such as model-specific and model-agnostic explanations.
\end{itemize}\end{minipage}\tabularnewline
\end{tabular}\\\\\\
Danilevsky et al. \cite{danilevsky2020survey}  &\begin{tabular}{l}
\begin{minipage}[t]{0.8\columnwidth}  The authors address  some common method levels of XAI such as explanation level, and implementation level. \end{minipage}\tabularnewline
\end{tabular}& \begin{tabular}{l}
\begin{minipage}[t]{0.8\columnwidth} \begin{itemize}
\item Our classification scheme is more systematic, multi-level and based on clear criteria at each level.
\item No sub-classification is provided for the studied challenges based on modular specific sub-challenges.
\item This survey is mainly dedicated to the Natural Language Processing (NLP) community, while our survey is dedicated to the security community. 
\end{itemize}\end{minipage}\tabularnewline
\end{tabular} \\\\\\
H.Li et al. \cite{li2020survey}  & \begin{tabular}{l}
\begin{minipage}[t]{0.8\columnwidth}  Similar to our survey, this survey takes into consideration some overlooked challenges in other surveys such as trust and reliance, and robustness. 
\end{minipage}\tabularnewline
\end{tabular}&  \begin{tabular}{l}
\begin{minipage}[t]{0.8\columnwidth} \begin{itemize}
\item In our work, we uncover several research directions that are not discussed in \cite{li2020survey}, including but not limited to: model dependence class in XAI which consists of model-specific and model-agnostic explanations.
\item This survey is mainly dedicated to the data and knowledge engineering community, while our survey is dedicated to the security community. 
\end{itemize}\end{minipage}\tabularnewline
\end{tabular}\\\\\\
Confalonier et al. \cite{confalonieri2021historical}  &  \begin{tabular}{l}
\begin{minipage}[t]{0.8\columnwidth} 
Similar to our survey, this survey discusses the method XAI  levels and presents a historical perspective of XAI.
\end{minipage}\tabularnewline
\end{tabular}& \begin{tabular}{l}
\begin{minipage}[t]{0.8\columnwidth} \begin{itemize}
\item No classification is provided for the techniques proposed to address the discussed challenges.
\item Our work addresses challenges of XAI  that are not discussed in \cite{confalonieri2021historical} such as the level of explanation, implementation level, and model dependence.
\end{itemize}\end{minipage}\tabularnewline
\end{tabular}\\\\\\
\end{tabular}
\end{table*}

\begin{table*}
\thisfloatpagestyle{empty}
\centering
\begin{tabular}{c c c }
\hline
Approach & Similarities & Differences\\ 
\hline \\

Mohseni et al. \cite{mohseni2021multidisciplinary}  &\begin{tabular}{l}
\begin{minipage}[t]{0.8\columnwidth}  Similar to our survey, this survey takes into consideration some overlooked challenges in other surveys such as trust and reliance. \end{minipage}\tabularnewline
\end{tabular}& \begin{tabular}{l}
\begin{minipage}[t]{0.8\columnwidth} \begin{itemize}
\item The classification scheme presented in this survey is based on the  XAI design goals and evaluation methods, resulting in three categories: desired properties, desired outcomes,  and practical approaches.
\item Our classification scheme is more systematic, multi-level and based on clear criteria at each level.
\item Our work addresses challenges of XAI that are not discussed in \cite{mohseni2021multidisciplinary} such as interactivity, stability and robustness.
\end{itemize}\end{minipage}\tabularnewline
\end{tabular}\\\\

Linardatos et al. \cite{linardatos2020explainable}  & \begin{tabular}{l}
\begin{minipage}[t]{0.8\columnwidth} Similar to our survey, this survey discusses the XAI method levels based on the level of explanation, implementation level, and model
dependence. \end{minipage}\tabularnewline
\end{tabular}& \begin{tabular}{l}
\begin{minipage}[t]{0.8\columnwidth} \begin{itemize}
\item  No sub-classification is provided for the studied challenges based on modular specific sub-challenges. 
\item In our work, we uncover several research directions that are not discussed in \cite{linardatos2020explainable}, including but not limited to some common mathematical formulation of common or popular XAI methods.
\end{itemize}\end{minipage}\tabularnewline
\end{tabular}\\\\\\
P.Angelov et al. \cite{angelov2021explainable}  & \begin{tabular}{l}
\begin{minipage}[t]{0.8\columnwidth} Similar to our survey, this survey discusses the XAI method
 levels and presents a historical perspective of XAI. \end{minipage}\tabularnewline
\end{tabular} & \begin{tabular}{l}
\begin{minipage}[t]{0.8\columnwidth} \begin{itemize}
\item  No classification is provided for the techniques proposed to address the discussed challenges. 
\item Our classification scheme is more systematic, multi-level and based on clear criteria at each level.
\end{itemize}\end{minipage}\tabularnewline
\end{tabular} \\\\\\

\hline\\
\hline\\
\end{tabular}
\end{table*}

\section{Classification of Scientific Articles on Explainability of AI Models for Cybersecurity} \label{Papers_Classification}

As a result of conducting a thorough analysis of all selected articles following the methodology explained in Section \ref{Methodology}, the following cybersecurity categories were identified:

\begin{itemize}
    \item \textbf{Intrusion Detection} \newline
    Intrusion Detection is the process of detecting unauthorized access or malicious activity on a computer network. This can be accomplished through the use of software that monitors network traffic and logs to identify potential security threats. Intrusion detection can help prevent data theft, unauthorized access, and other network security breaches, and is an important component of a comprehensive security strategy.\\
    \item \textbf{Intrusion Prevention}\newline
    Intrusion Prevention is a proactive approach to network security that aims to stop malicious attacks and unauthorized access attempts before they can cause harm to a  network. This is done by using a combination of technologies, such as firewalls, and other network security tools that are designed to identify and prevent security threats in real-time. Intrusion prevention systems can block suspicious network traffic, prevent malicious code from executing, and alert security personnel to potential security incidents. The goal of network intrusion prevention is to reduce the risk of network security breaches and to minimize the impact of attacks that do occur.\\
    \item \textbf{Access Control}\newline
    Access control is a security measure that regulates who or what is allowed to access a computer network, and what actions they can perform. The goal of access control is to ensure that sensitive information is protected, and that only authorized users are able to access it.
There are several methods of implementing access control, including:
\begin{enumerate}
\item Identification and authentication: This involves verifying the identity of a user before granting access. This is usually done through the use of usernames and passwords, but can also involve other forms of authentication such as smart cards, biometrics, or security tokens.
\item Authorization: This involves granting specific permissions to users based on their role or job function. For example, a user may be allowed to view certain files but not allowed to modify them.
\item Access control lists (ACLs): This involves creating a list of permissions for each user of the network, specifying what they are allowed to access and what actions they can perform.
\item Role-based access control (RBAC): This involves assigning users to roles within an organization network, and then granting access based on those roles.
Access control is an important component of a comprehensive network security strategy, as it helps prevent unauthorized access to sensitive information and reduces the risk of security breaches.\\
\end{enumerate}
    \item \textbf{Privacy}\newline
    Privacy is important to maintain the security of sensitive information and to prevent identity theft, financial fraud, and other malicious activities through a network. It is also essential to building trust in technology and ensuring that individuals have control over their personal information and data.\\
    \item \textbf{Authentication}\newline
    Authentication in cybersecurity refers to the process of verifying the identity of a user, device, or system before granting access to a network system. The goal of authentication is to ensure that only authorized individuals and devices have access to sensitive information and resources.
There are several methods of authentication, including:
\begin{enumerate}
\item Passwords: The most common form of authentication, where a user provides a secret string of characters to prove their identity.
\item Multi-factor authentication (MFA): A method of authentication that requires more than one form of proof of identity. This can include a combination of passwords, security tokens, biometrics, and smart cards.
\item Biometrics: The use of physical characteristics, such as fingerprints, facial recognition, or iris scans, to prove identity.
\item Security tokens: Physical devices that generate one-time codes, which are used to authenticate a network user or device.
\item Digital certificates: Digital credentials that are issued by a trusted authority and used to authenticate the identity of a network user or device.
Authentication is an important component of cybersecurity as it helps prevent unauthorized access to sensitive information and resources through a network system. By requiring users to prove their identity before granting access, organizations can reduce the risk of security breaches and protect against malicious attacks.
\end{enumerate}
    \item \textbf{Trust and Reputation}\newline
    Trust and reputation in cybersecurity refer to the confidence that individuals and organizations have in the security and reliability of online systems, devices, and services. Trust and reputation are important components of cybersecurity because they help to ensure that users have confidence in the security and privacy of their online information, and that network systems and devices are secure and reliable.
The reputation of a network system, device, or service can be determined by several factors, including:
\begin{enumerate}
\item Past security incidents or breaches.
\item The level of security and privacy measures in place.
\item The level of transparency and accountability of the network system or service provider.
\item User feedback and recommendations.
\item  Independent security evaluations and certifications.
\end{enumerate}
Building trust and maintaining a positive reputation are important for organizations and individuals as they help to build trust and confidence among users and stakeholders, and to prevent network security breaches and other malicious activities. To enhance trust and reputation in cybersecurity, organizations should focus on implementing robust network security and privacy measures, maintaining transparency and accountability, and responding promptly to network security incidents and online data breaches.
\end{itemize}

We will discuss the-state-of-the-art for each one of the six  cybersecurity classes in Section \ref{xai-cyb} and identify a set of criteria that we believe important to consider when designing future XAI solutions for cybersecurity.
\begin{figure*}[!ht]
 \includegraphics[width=\textwidth,height=.7\textwidth]{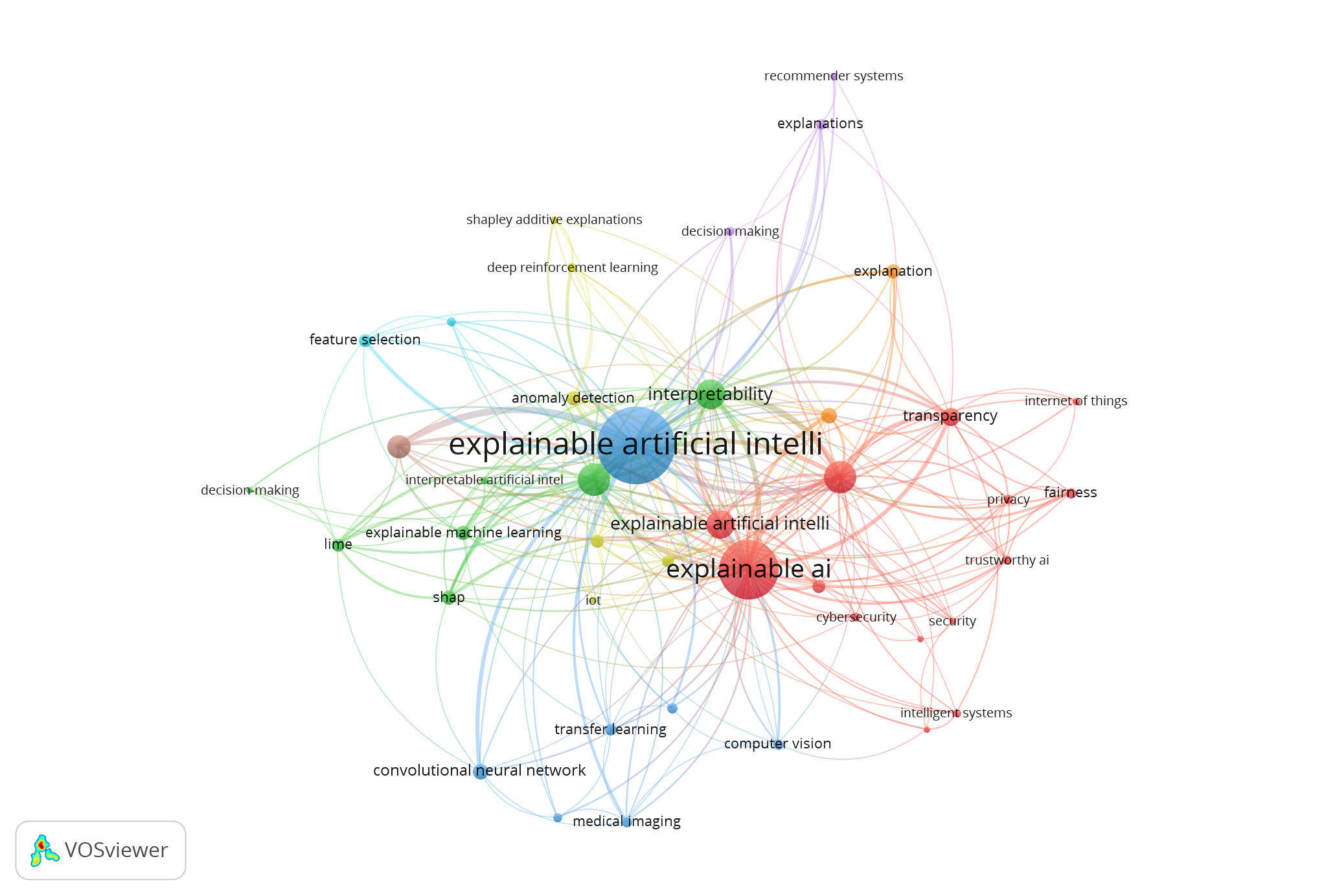}
  \caption{The map of author keywords for XAI publications}
  \label{fig5}
\end{figure*}

In order to gain a full picture of the research area of XAI in  cybersecurity, we used bibliometric analysis. Our study shows the state of the field by presenting a map of the main problem-specific notions based upon author keywords, as well as an annual scientific production report.  We utilized the tool VOSViewer \cite{van2013vosviewer} to generate the author keyword maps. 
\begin{figure*}[!htbp]
 \includegraphics[width=\textwidth,height=.7\textwidth]{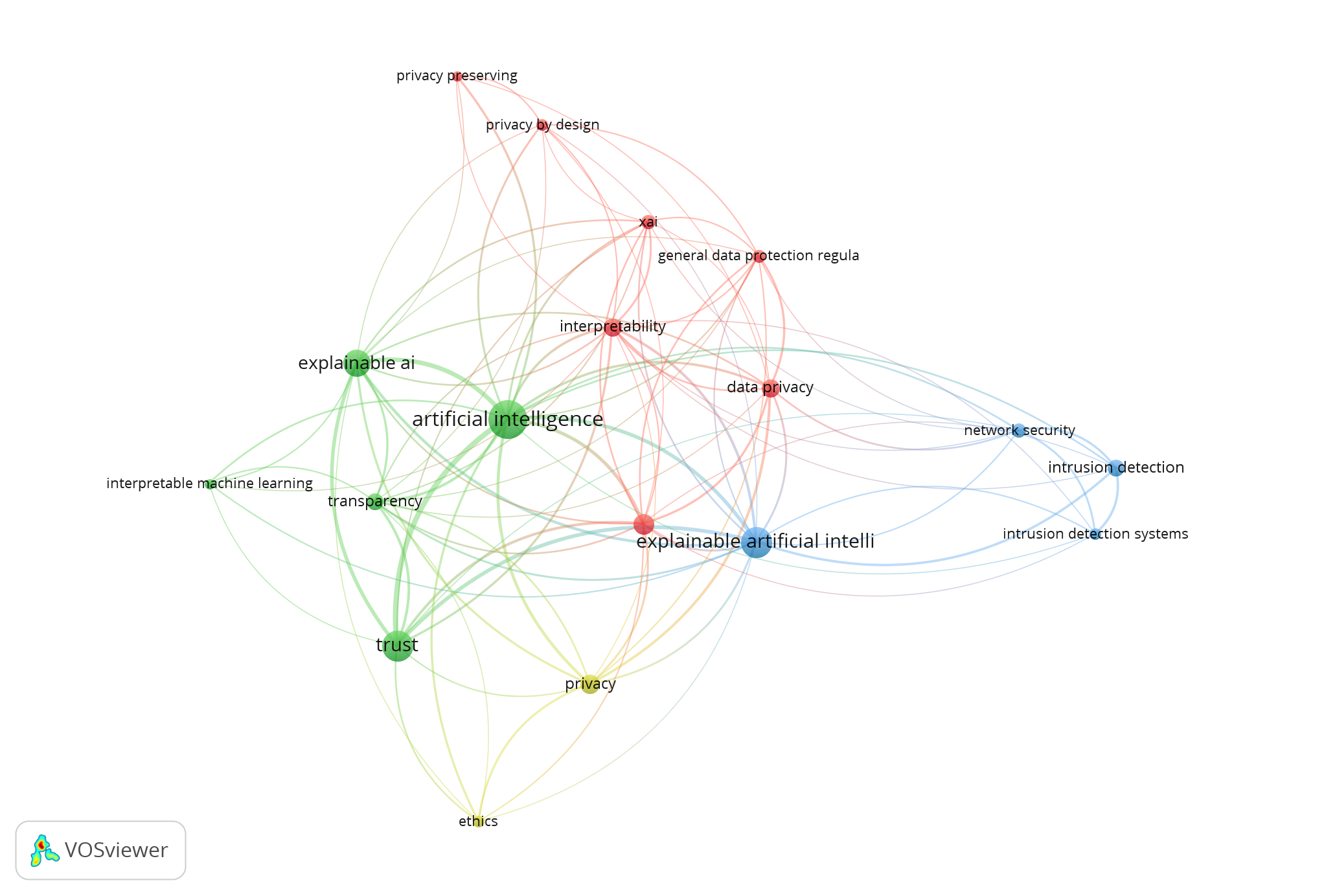}
  \caption{The map of author keywords for XAI with cybersecurity publications}
  \label{fig6}
\end{figure*}
Fig. \ref{fig5} shows a graph containing the most popular author keywords for XAI. It can be concluded that XAI is often investigated in the context of AI (i.e.  ‘‘NLP’’, ‘‘CNN’’, "deep reinforcement learning", etc) as well as medical imaging (i.e. ‘‘COVID-19’’, ‘‘clinical support system’’, etc.) and cybersecurity (i.e. ‘‘trust’’, ‘‘anomaly detection’’, ‘privacy’’, etc). 
Similar notions are observed to be essential for cybersecurity  explanation as shown in Fig. \ref{fig6}. However, a distinction between different clusters in the latter case is visible more clearly. This is hypothesized to be due to a more diverse usage of the term XAI across various cybersecurity areas.

\begin{figure*}[!ht]
 \includegraphics[width=\textwidth,height=.4\textwidth]{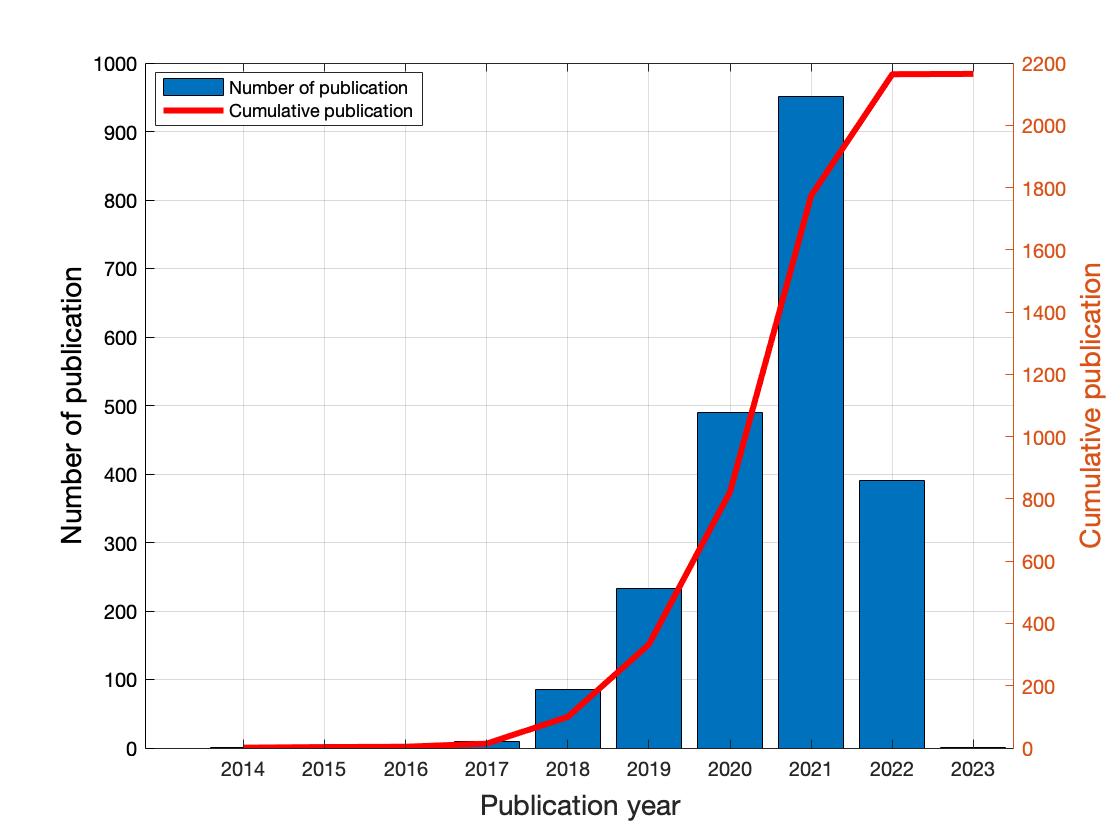}
 \caption{Annual scientific production for XAI publications}
  \label{fig3}
\end{figure*}

\begin{figure*}[!ht]
 \includegraphics[width=\textwidth,height=.4\textwidth]{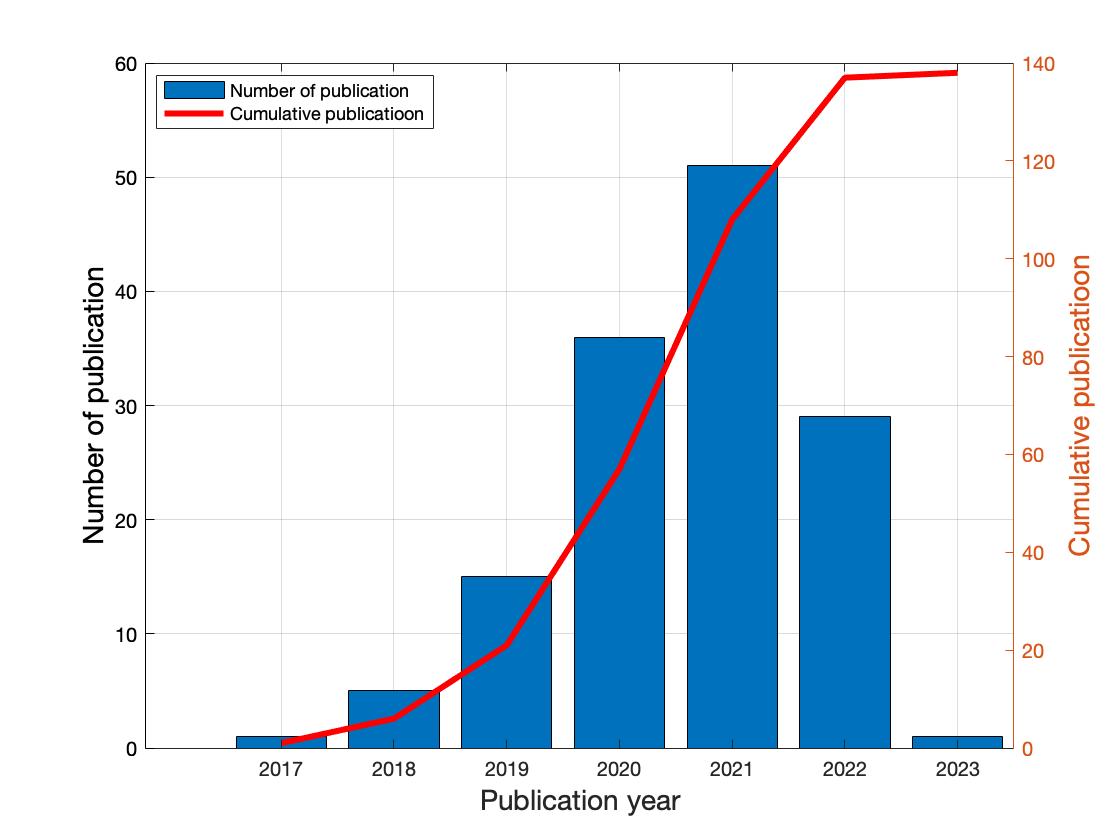}
  \caption{Annual scientific production for XAI with cybersecurity publications}
  \label{fig4}
\end{figure*}

XAI has emerged as a key area of research in recent years, attracting attention across a wide range of subject areas. As shown in Fig. \ref{fig3}, there has been a significant increase in the number of XAI publications over the past three years, indicating the growing interest in this field. In particular, we focus on the intersection of XAI and cybersecurity, as shown in Fig. \ref{fig4}. It should be noted that the number of publications in 2023 is limited by the search date, and the actual number of publications is likely to be higher.

Although the categories we identified in this section are carefully designed to be comprehensive, it is important to recognize their limitations and weaknesses that need to be considered when conducting further research or implementing solutions. One of the main limitations of these categories is that they may not be exhaustive or applicable to all types of AI models or cybersecurity scenarios. Therefore, it is crucial to approach the analysis with a critical perspective and to assess the categories' relevance and usefulness in each specific context. Additionally, the categories may not reflect the most recent developments or advancements in the field. This is due to the fact that the field is constantly evolving and expanding at a rapid pace, with new technologies and methodologies being introduced regularly. Another limitation of the identified categories is that they may not fully capture the complexity of the AI model explainability problem in cybersecurity. AI models can be intricate and difficult to interpret, and their behavior may change over time or under different conditions. Therefore, it might be challenging to create a comprehensive set of categories that cover all possible scenarios and situations.

\section{Explainability via Mathematical Models} \label{Math_Models}
This section introduces the concept of XAI along with its mathematical concepts. Afterwards, we discuss the main similarities and differences between XAI and other related AI concepts such as interpretability, transparency, and robustness. Finally, we explain how XAI is related to the substantial cybersecurity concepts, i.e., trust, intrusion detection, authentication, intrusion prevention, access control, and privacy.

As shown in Table \ref{table2}, all mathematical equations and algorithms are explained using a common set of notations. These mathematical equations may differ from those published in their respective research publications since we are keen on ensuring the consistency of the notations that we use throughout the survey. The aim is to facilitate the paper's reading and establish a common repository of notations.
We start by explaining the main XAI techniques in the literature, i.e., Shapley value (SHAP), Local Interpretable Model-Agnostic Explanations (LIME) and and Layer-wise Relevance Propagation (LRP).

\begin{table}[h]
\caption{Table of notations}
\label{table2}
\begin{tabular}{|c|p{6cm}|}
    \hline
    Notation & Description\\
    \hline
    $i$ & A statistical unit 
    \\
    \hline
    $X_i$ & A vector of explanatory attributes, where $x_{ik}$ is the feature variable $k$ of $X_i$ \\
    \hline
    $k$ & A single feature variable from the total number of explanatory variables $K$  \\
    \hline
    $\theta_{k}(X_i)$ & The local functions that  represent the Shapley values\\
    \hline
    $S_i$ & The feature subset that are used to retrain the model
    \\
    \hline
    $ \hat{f} \left( S_i\right)$ & The predictions associated with all the possible model configurations\\
    \hline
     $ G$ & A set of interpretable models, where $g \in G$ is a typical model\\
    \hline
    $ \Omega(g)$ & A measure of complexity of the explanation $g \in G$\\
    \hline
     $ \xi(x)$ & The explanation produced by LIME\\
    \hline
  $\pi_x$ & proximity measure, which defines the size of the neighborhood around the instance $x$\\
    \hline
   $L (f , g, \pi_x )$ & A measure of how unfaithful $g$ is in approximating $f$ in the locality defined by $\pi_x$\\
    \hline
 $z^{'}$ & The perturbed feature of selected instances  to recover the original representation for selected instances $z \in \mathbb{R}^{d}$\\
    \hline
    $R_{m}^{l}$ & The relevance score of the $m^{th}$ neuron in layer $l$\\
    \hline
      $z_{jh}$ & The quantity contribution made by neuron $j$ to the relevance of neuron $h$\\
    \hline
\end{tabular}
\end{table}

\subsection{Shapley Value (SHAP)}
 In explainable ML, the Shapley value is one of the most commonly used methods to represent the importance placed on an input feature by a model. SHAP \cite{lundberg2017unified} is a method used in explainable machine learning (ML) to assign a value to each feature, or input, that represents its importance to the model. It is based on the Shapley value from cooperative game theory and is a way to fairly distribute a value among a group of individuals, in this case, the input features. The Shapley value is commonly used in explainable ML because it provides a way to understand how a model is making its predictions by quantifying the contribution of each feature to the overall prediction. It can be used for both global and local explanations, and it is unique in the sense that it satisfies several desirable properties, such as consistency, efficiency, and locally accuracy.
 In recent years, it has received increasing attention in the AI community and has been applied in a wide range of domains, including classification, discovery, and automation.  
The Shapley value measures an input feature’s influence on the model’s output score, according to its definition, by the average influence that it has on the model’s output score, across all the possible classifications. The more the Shapley value increases, the more the model’s predictions assign importance to the input feature.

Generally, Shapley value satisfies a variety of helpful criteria that facilitate a better understanding of how the model employs its features to provide a trustworthy answer in a complex decision-making process. In a linear model, the Shapley value of a feature is the linear combination of its Shapley values across the model. For example, the sum of the Shapley values equals the accuracy of the model; they are the same for features of equal significance.

Let $i = 1,\dots, n$ be a statistical unit and $Y_i$ be multivariate features to be predicted on a test set such that an automated action $a(Y_i)$ is performed. 
Using a vector of explanatory attributes $X_i$ and a ML model $l$, we can get the predicted value for the response vector ${\hat{Y}_i}^l$ as a function of $X_i$: ${\hat{Y}_i}^l = {\hat{f}}^l (X_i)$. As a matter of convenience, the suffix $l$ will no longer be used. 

A function $\Theta$ may be used to break a ML model into functions of the additional individual components $x_{ik}$ of the features vector $X_i$. In the case of a linear model that belongs to the class of additive variable attributions, the function $\Theta$ can be given as follows:

\begin{equation} \label{Shap.}
\Theta \left( \hat{f} \left( X_i \right) \right) = \theta_0 + \sum_{k=1}^{K} \theta_k x_{ik}, \quad \forall i=1,\dots ,n.
\end{equation}

\noindent where $\theta_0$ is the intercept, $K$ is the total number of explanatory variables,  the coefficient vector $\theta  \in \mathbb{R}^{k}$ indicates the orientation of the global model plane with constant slope in each direction of the input space, and $n$ represents the total number of units to be predicted. 
In fact, the local functions $\theta_{k}(X_i)$ are the Shapley values.

Thus, as shown by Eq. \ref{Shap.}, the authors in \cite{giudici2021shapley,aas2021explaining,bussmann2021explainable} suggest to get a linear approximation of a ML model by regressing the response values on the individual Shapley values. Despite the fact that this proposal gives statistically testable local explanations, it may lead to a heavily parameterized model that is computationally demanding. This is due to the fact that, as in a regular model selection approach, the equation approximation must be examined for potential subsets of the $K$  variables.
In the context of a ML model, the $K$ explanatory variables that can be included in the model can be thought of as the participants in a cooperative game with the goal of producing a payoff. Each model is a combination of several variables, which cooperate towards making predictions $\hat{f} \left( X_i \right)$.

A feature significance strategy for linear models in the presence of multicollinearity is proposed by the authors of \cite{lundberg2017unified,basu2022multicollinearity} and as a Shapley regression values method. All feature subsets $S_i\subseteq X_i, \forall i \in \{1,\dots,n\}$ are used to retrain the model. 
Each feature is given a value that reflects how important it is to the model's prediction. By training two models, one with and one without a particular feature, the impact of the feature is estimated by comparing the performance of the two models. Then, predictions from the two models are compared on the current input $\hat{f} \left( S_i \cup \{x_{ik}\} \right) -  \hat{f} \left( S_i\right)$. 
All alternative subsets of $S_i\subseteq X_i \setminus \{x_{ik}\}$ are taken into account when determining whether or not withheld a feature has any influence on the model, where $X_i \setminus \{x_{ik}\}$ is the set of all the possible model configurations which can be obtained with $K-1$ variables excluding the variable $x_{ik}$. Once the Shapley values have been calculated, they are utilized as feature attributions. A weighted average of all potential variations is computed as follows:

\begin{equation}
\begin{aligned}
\theta \left( \hat{f} \left( X_i \right) \right) = \sum_{S_i \subseteq X_i \setminus \{x_{ik}\}} \frac{\mid S_i \mid ! (K - \mid S_i \mid -1 )! }{K!}\\ [ \, \hat{f} \left( S_i \cup \{x_{ik}\} \right) - \hat{f} \left( S_i\right) ] \,
\end{aligned}
\end{equation}

\subsection{Local Interpretable Model-Agnostic Explanations (LIME)}

The LIME method is a model-agnostic, ML-based approach to explain predictions or decisions. LIME was introduced by \cite{ribeiro2016should} with the goal of providing a model-agnostic explanation that enables users to understand the decision or prediction that a ML model made. LIME uses a variety of ML algorithms, including neural networks, deep learning, and reinforcement learning. It then uses a variety of methods to explain the decision or prediction, including abstract, intuitive, and stylized explanations. The LIME method has been used in a variety of domains, including health care \cite{loveleen2022explanation, kamal2022explainable}, agriculture \cite{viana2021evaluation}, and cybersecurity \cite{khan2022xsru}. 

Let's define the explanation  as a model $g \in G$, where $G$ is a set of interpretable models, such as decision trees, linear models,  or falling rule lists. Let  $\Omega(g)$ be a measure of complexity of the explanation $g \in G$. To have a simple model that can be interpretable, low $\Omega(g)$ should be the goal, for example, for decision trees $\Omega(g)$ may be the depth of the tree where the shortest depth is more interpretable. 

The model being explained is denoted by $f : \mathbb{R}^d \rightarrow  \mathbb{R}$ where $d$ is the model's dimension. The explanation produced by LIME is obtained by the following minimization problem:

\begin{equation}
\xi(x) = \arg\min_{g \in G} L(f,g,\pi_x) + \Omega(g)
\end{equation}

\noindent where $\pi_x$ is the proximity measure to $x$, which defines the size of the neighborhood around the instance $x$, and $L (f , g, \pi_x )$ is a measure of how unfaithful $g$ is in approximating $f$ in the locality defined by $\pi_x$. The final goal is to minimize $L (f , g, \pi_x )$ (defined in Eq. \ref{Lime}) and to get an interpretable approximation of the black-box model.

\begin{equation}\label{Lime}
 L(f,g,\pi_x) = \sum_{z,z^{'} \in Z} \pi_{x}(z) ((f(z) - g(z^{'}))^2
\end{equation}

\noindent where $\pi_{x}(z)$ as a proximity measure between an instance $z$ and selected instance $x$, so as to define locality around $x$.
  $\pi_{x}(z)$ is given as follows: $\pi_{x}(z) = exp (-D(x,z)^{2} / \sigma ^2)$, which gives the weighted probability of other instances being sampled, where  $D$ can be any distance metrics between $x$ and $z$ (the feature of other instances), and  $\sigma $ is a hyperparameter for width.
Finally, $z^{'} \in \{0,1\}^{d^{'}}$ is the perturbed feature of selected instances  to recover the original representation for selected instances $z \in \mathbb{R}^{d}$.

\subsection{Layer-wise Relevance Propagation (LRP)}

LRP is one of the more promising methods of XAI. It encapsulates the idea of relevance propagation in an elegant way and has achieved impressive results. It has become the basis for many popular contextual XAI systems. LRP provides an explanation for a classifier's prediction that is specific to a given data point by assigning relevance values to significant components of the input. This is accomplished through the utilization of the topology of the trained model itself.
Specifically, LRP aims to explain neural network predictions by indicating what aspect of the input data supports the actual prediction. It has been effectively applied to text, images, and videos, where the output predicted value is used to compute the relevance value for the neurons in the lower layer of the neural network. The importance of a neuron in the backward pass increases with its forward pass effect increases. The neurons with the highest backward pass effect are the least relevant.This is a generic technique for learning relevance propagation. The input to the network includes only the relevant features; the output is the new set of features that are relevant to the input. If the input is highly relevant, the output will be highly relevant as well. LRP has been applied to many applications, including computer vision \cite{huang2021visual,sun2022explain}, natural language processing (NLP) \cite{gholizadeh2021model}, recommendation \cite{nguyen2021quantifying}, object recognition \cite{achtibat2022towards}, and anomaly detection \cite{ide2021anomaly}.

LRP assigns relevance scores, which indicate a neuron's contribution to the network's output, to all neurons in each neural network layer via backpropagation.  Upper-layer neurons redistribut their relevance scores to the immediate lower-layer neurons. In order to meet the conservation property, as demonstrated in Eq. \ref{eq6} below, the total of the relevance scores in each layer must equal the output. 

\begin{equation}
\sum_{m} R_{m}^{l} = R_{j}^{l-1} = R_{h}^{1}
 \label{eq6}
\end{equation}

\noindent where $R_{m}^{l}$ is the relevance score of the $m^{th}$ neuron in layer $l$, and $j$ and $h$ are neurons at two consecutive layers of the neural network. The following rule is used to propagate relevance scores $(R_{h})$ from a given layer onto neurons in the lower layer: 

\begin{equation}
R_j  = \sum_{h} \frac{z_{jh}}{\sum_{j} z_{jh}} R_h
 \label{eq7}
\end{equation}

\noindent where $z_{jh}$ represents the quantity contribution made by neuron $j$ to the relevance of neuron $h$. Once the input features are reached, the propagation process terminates. 

In this section, while we provide the solid foundation of XAI in mathematical underpinnings, it is important to acknowledge the limitations of XAI techniques and to explore their relationship with cybersecurity concepts in greater detail. One key limitation of XAI techniques for cybersecurity is their interpretability-accuracy trade-off, which often requires sacrificing accuracy for interpretability. Additionally, some XAI techniques may be biased or fail to capture complex interactions between relevant cybersecurity variables, leading to inaccurate or misleading explanations with regard to false positive and false negative attacks. Thus, it is important to carefully evaluate the suitability of XAI techniques for specific cybersecurity use cases and to develop approaches that balance accuracy and interpretability. Furthermore, a more in-depth exploration of the relationship between XAI and cybersecurity concepts could yield valuable insights into how XAI can be applied to enhance cybersecurity defense and response strategies. For example, XAI techniques could be used to identify and mitigate potential attacks, improve threat intelligence, and enhance user understanding of system behavior. Ultimately, a critical and nuanced understanding of XAI's strengths and limitations is essential to effectively apply these techniques in the field of cybersecurity.

\section{XAI Methods Classification}
\label{xai-methods}
\begin{figure*}[!ht]
 \includegraphics[width=\textwidth,height=.5\textwidth]{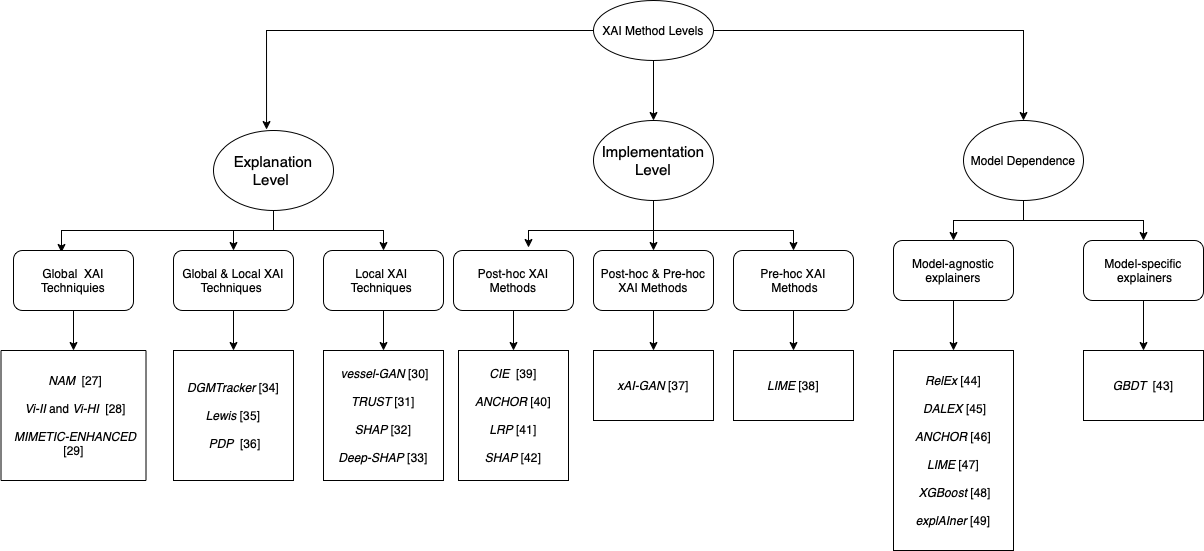}
  \caption{Classification of the XAI papers based on the high-level challenge that they address}
  \label{fig77}
\end{figure*}

The XAI methods can be classified based on the level of explanation, implementation level, and model dependence as shown in Fig. \ref{fig77}. An XAI technique that focuses on the whole model or just a single instance is called an explanation level. The subcategories of the explanation level of an XAI method are named global and local, respectively. Global XAI techniques explain the whole model and are most often used to validate the predictions of other models. Local XAI techniques explain only a single instance of the model and are most often used to evaluate the performance of a particular decision or action. Some XAI methods, such as Neural Additive Model (NAM) in \cite{utkin2022extension}, explainable Deep Belief Networks (DBN) with Visual Input-neuron Importance (\textit{Vi-II}) and Visual Hidden-layer Importance (\textit{Vi-HI}) in \cite{chen2021novel}, and an enhanced MultImodal deep learning-based MobilE TraffIc Classification  (\textit{MIMETIC-ENHANCED}), an architecture for Traffic Classification (TC) operating at biflow level in \cite{nascita2021xai} used global XAI techniques, but some employed only local XAI techniques, such as the explainable Generative Adversarial Networks (GAN) named \textit{vessel-GAN} in \cite{wu2022vessel}, Transparency Relying Upon Statistical Theory (\textit{TRUST}) 
\cite{zolanvari2021trust}, explaining anomalies detected by autoencoders using SHAP \cite{antwarg2021explaining}, and explainable Deep Reinforcement Learning (DRL) based on Shapley additive named the \textit{Deep-SHAP} method \cite{zhang2021explainable}. Other XAI techniques used global and local  techniques together, such as  \textit{DGMTracker} that explains Deep Generative Models \cite{liu2017analyzing}, \textit{Lewis} \cite{wang2021demonstration}, and Partial Dependence Plots (\textit{PDP}) \cite{peng2021explainable}.

The implementation level of a XAI method is classified into two classes: Post-hoc explanation methods and Pre-hoc explanation methods. Post-hoc explanation methods are used to explain the performance of a model after it has been trained. Pre-hoc explanation methods are used to explain the performance of a model before it has been trained. Some XAI techniques used both pre-hoc and post-hoc explanation methods, such as the case of \textit{xAI-GAN} \cite{nagisetty2020xai} that employed saliency map, LIME, and DeepSHAP, which provide “richer” feedback from the discriminator to the generator to enable more guided training and greater control. Some other XAI techniques used only pre-hoc explanation methods, such as the case of  \cite{kamal2021alzheimer}, which explains a model's predictions using \textit{LIME}. Other XAI techniques employed only post-hoc explanation methods, such as in the case of Confident Itemsets Explanation (\textit{CIE}) \cite{moradi2021post}, which explains a model's predictions using the language model trained on a large corpus of text, \textit{ANCHOR} \cite{nourani2021anchoring}, \textit{LRP} \cite{lee2021explainable}, and \textit{SHAP} \cite{knapivc2021explainable}.

Model dependency class in XAI consists of model-specific and model-agnostic explainers.  Model-specific explainers are designed to provide explanations for a specific model from a specific domain or domain-specific knowledge, such as a Gradient Boost Decision Trees (\textit{GBDT}) \cite{delgado2022implementing}, by making predictions for every possible example in a dataset. Model-agnostic explainers are designed to provide explanations for any model, no matter the domain or domain-specific knowledge. Since most of the model-agnostic explainers also provide post-hoc explanations, these methods have been often used due to their flexibility. \textit{RelEx} \cite{zhang2021relex}, \textit{DALEX} \cite{srinath2022explainable}, \textit{ANCHOR} \cite{prentzas2021model}, \textit{LIME} \cite{magesh2020explainable}, \textit{XGBoost} \cite{ahmed2021explainable}, and \textit{explAIner} \cite{spinner2019explainer} are all examples of model-agnostic explainers.

\section{XAI-based Cybersecurity}
\label{xai-cyb}

As discussed earlier in Section \ref{related}, very few surveys have addressed the application of XAI to cybersecurity, but no paper has focused on the network and system security subfield. It is important, though, to draw attention to this topic, particularly since the use of technologies such as big data, cloud computing, and IoT is continually growing, along with the need to employ technologies (i.e. AI and ML) that prevent attacks on data and networks.
To this end, network and system security can benefit greatly from XAI and the advancements in ML to avoid the mass exploitation of vulnerabilities that can be used by attackers for personal gain and to implement measures that can detect and remediate advanced  cyberattacks. Considering the above, cybersecurity linked to XAI, particularly the use of AI/ML for its implementation, can help users avoid attacks that are easy to launch but difficult to detect and can block access to critical systems as a means of carrying them out. Moreover, AI and ML techniques applied to cybersecurity can offer a better approach for detecting cyberattacks, such as botnets, brute force attacks, or attempts to penetrate a network system with weak authentication.

AI, in particular the field of ML, has also gained attraction recently in the cybersecurity field, with large number of works on automated diagnosis, prognosis, and risk assessment. Recent research has also focused on using AI and ML for incident response, in order to quickly detect and mitigate new threats \cite{kuppa2021adversarial}. This is especially critical as the cyber threat landscape changes rapidly, with zero-day exploits becoming a rarity, and more traditional attack vectors like web applications and email becoming the preferred method of an attacker. The AI techniques used in these areas are also finding applications in other industries, such as finance, advertising, and healthcare. For instance, AI is being used to analyse large amounts of data in order to discover trends and correlations that human analysts would otherwise miss. This enables the discovery of fraudulent or anomalous transactions that would have gone unnoticed by human analysts.   
To pay more attention to this topic and consider how it fits into XAI, we will assess the results of the proposals that have been launched on these topics. We will perform this assessment in light of our cybersecurity classification we present in the following.

\subsection{Privacy} 
XAI can play a significant role in ensuring privacy by providing a deeper understanding of how AI models handle and process sensitive information related to cybersecurity threats and vulnerabilities. For example, XAI can help us identify any potential privacy issues that may arise in the data processing and analysis stages, such as the unauthorized access or sharing of sensitive information. Additionally, XAI can provide insights into the model's decision-making process and help us ensure that the data is being used for its intended purpose, such as detecting and preventing cyber attacks.
In order to increase privacy in XAI for cybersecurity, researchers  explored various techniques and approaches, such as differential privacy \cite{saifullah2022privacy,wang2022consensus}, federated learning \cite{wahab2021federated,rjoub2022trust22,huong2022federated,barcena2022approach}, and homomorphic encryption \cite{muftuouglu2022privacy}. These techniques aim to protect sensitive data while still allowing the model to learn and make accurate predictions about cyber threats and vulnerabilities.

Many studies mentioned privacy in the context of explainability, since an explanation can sometimes be used to gain a deeper understanding of a model \cite{milli2019model,tramer2016stealing}.  Understanding the model this way will help explain how the model works and the relationships between the variables that the model is based on. On the other hand, used to learn about the training data \cite{shokri2019privacy}, an explanation can land a hand to the user to identify trends and patterns, which is highly useful for making decisions, for example, on how to avoid overfitting and increase accuracy. For example, the authors in \cite{harder2020interpretable}, develop  an explanation model (LEM) that uses a collection of locally linear maps to approximate complex models and provide interpretable and differentially private explanations for classification decisions. The method was shown to achieve high classification accuracy on several image benchmark datasets.
LEM works by partitioning the input space into a set of locally linear regions, and fitting a linear classifier to each region using a subset of the original training data. This results in a collection of local classifiers that can be used to make predictions on new data points. To ensure privacy, the local classifiers are perturbed using the Differential Privacy (DP) framework, which adds noise to the local model parameters.

In \cite{lobner2021explainable}, the authors propose an explainable model for predicting default privacy settings. Compared to previous work, they introduce several improvements in the model design, evaluation, and interpretability.
One key improvement is the use of an improved feature selection process that takes into account the importance of each feature in the prediction task. This helps to improve the accuracy and interpretability of the model. Additionally, the authors introduce enhanced evaluation metrics that can identify weaknesses in the model's design before it goes into production, improving the robustness and reliability of the model.
Another important contribution of the paper is the emphasis on interpretability and transparency in the model design. The authors propose a step-by-step approach to building the model that makes it easy for users to understand how the model works and why it makes certain predictions. This can help users build trust and confidence in the model and increase the likelihood that they will adopt it.


\subsection{Trust and Reputation}
Trust is one of the most pressing cybersecurity challenges we face today. AI systems that highlight the benefits of transparency and the potential for systems and data misuse can increase the frequency and amplify the impact of attacks. Because automated systems perform so many complex decisions, they can be exploited to harm more people or cause more damage than a human could. For these reasons, trust plays an important role in the safety of the AI technology we use, and our ability to accept control over it by humans. On the other hand, many studies mentioned trust in the context of explainability. According to these studies, trustworthy is defined by a number of factors, including reliability \cite{rjoub2020trust,rjoub2022trust,drawel2022formal}, explanation of results \cite{machlev2021measuring}, respect for privacy \cite{kuppa2020black}, and inability to exhibit bias \cite{rjoub2022explainable,rjoub2022one,rjoub2022active}.
In \cite{zolanvari2021trust}, the authors presents a statistical model that aims to generate adequate trust on AI systems by introducing the Transparency Relying Upon Statistical Theory (TRUST) XAI model. TRUST XAI is model-agnostic, high-performing, and suitable for numerical applications. It simulates the statistical behavior of the outputs in an AI-based system. The input features are transformed into latent variables using factor analysis. In order to rank these variables, the authors use mutual information to pick just the most influential ones, which they call "representatives" of the classes. Multi-modal Gaussian distributions are then used to determine the likelihood of new samples belonging to each classification. Using three different  cybersecurity datasets, they demonstrate TRUST XAI's effectiveness in a case study on network security of the industrial Internet of things (IIoT). 
Researchers propose in \cite{mankodiya2021xai} an approach based on ML to detect misbehaving vehicles in a vehicular adhoc network (VANET). To classify the attackers, they use decision tree-based algorithms. As part of XAI, they provided insights on model performance by using evaluation measures for trust management.
In \cite{elayan2021internet}, the authors introduce the concept of Internet of Behavior (IoB) and its integration with XAI techniques to provide an unequivocal and trustworthy experience in order to change IoT behavior to improve user behavior. As a result, a system based on IoB and XAI has been developed in order to influence consumer behavior to reduce power consumption and costs with electrical power consumption as the use case scenario.

\subsection{Intrusion Detection}
Intrusion Detection (ID) in cybersecurity has become a critical component of any network security strategy. As a result, it has become a common target for cyber attackers looking to compromise the system \cite{wahab2022intrusion}. Advances in ID technologies and risk-based management have made it more difficult to breach the system. 
XAI adds another layer of protection to an organization’s cybersecurity team. This helps discover new threats quickly and take action against them.  XAI can also be used to spot when a potential intruder is in the system. In particular,  XAI can be used to identify potential intrusions in a system by providing insight into the decision-making processes and behaviors of the system. This can be particularly useful in detecting anomalies or unusual patterns that may indicate the presence of an intruder.
One approach to using AI for network intrusion detection is to train a machine learning model on normal system behavior and use it to identify and explain deviations from this behavior. For example, an AI system might be trained to recognize patterns of system usage that are typical for a particular user or group of users, and then use this knowledge to identify deviations from these patterns that could indicate an intruder.
Another approach is to use AI to analyze the output of an IDS and provide a human-understandable explanation of the system's decision-making process. This can help security analysts understand the basis for an alert and determine whether it is a genuine threat or a false positive.
Overall, AI has the potential to improve the effectiveness of intrusion detection systems by providing greater transparency and insight into their decision-making processes, which can typically assist in identifying and responding to potential threats more quickly and accurately. 

AI does not replace network security guards, but it can assist them identify potential breaches earlier and faster, and work together to protect the system.
However, one main limitation of using AI for intrusion detection is that most existing AI-based solutions lack explainability. This makes them unable to provide sufficient evidences about the  methodology for identifying cyberattacks as such. For example, the features that lead to identify a certain cyberattack must be clearly interpreted along with the decision. In the following, we will discuss the main approaches that use XAI for network intrusion detection. 

Based on monitoring Linux-kernel system calls (syscalls), the researchers in \cite{karn2020cryptomining} present a ML-based detection system of anomalous pods in a Kubernetes cluster. A number of ML models are built to detect anomalous pods among numerous healthy cloud workloads using cryptominers images as containers. SHAP, LIME, and an auto-encoding-based scheme for LSTM models are used to provide explanations. In \cite{wang2020explainable}, the authors present a methodology for improving the understanding of ID systems (IDSs) that leverages Shapley additive explanations that combines local and global explanations. The local explanations explain why the model makes particular decisions based on the input. The global explanations describe the links between the feature values and the various sorts of assaults and present the relevant characteristics retrieved from IDSs. Simultaneously, the interpretations of two distinct classifiers are compared: one-vs-all classifier and multiclass classifier. The authors in \cite{baryannis2019predicting} first introduce a supply chain risk prediction framework based on data-driven AI approaches and the synergy between AI and supply chain expertise. Then, they investigate the trade-off between prediction performance and interpretability by building and applying the framework to the issue of forecasting delivery delays in a real-world multi-tier manufacturing supply chain.
Given the power of deep learning algorithms to address real-world issues like anomaly detection and forensic investigation, the authors in \cite{aguilar2022towards} offer an interpretable autoencoder based on decision trees that is designed to handle categorical data without requiring data transformation. Furthermore, for specialists in the application domain, the interpretable autoencoder gives a natural explanation.

\subsection{Intrusion Prevention}
Intrusion prevention systems (IPSs) differ from detection systems in that they are focused on preventing the system from being compromised rather than detecting the actual attack after it happens. In other words, IPSs prevent an intruder from gaining access to the system and compromising it, while detection systems prevent an intrusion from being successful. XAI can assist network intrusion prevention systems (IPSs) by providing insight into the decision-making processes and behaviors of the network system. This can be particularly useful in helping security analysts understand why an IPS has identified a particular event as a potential threat, and in determining the appropriate response to that threat.
One way that XAI can assist IPSs is by providing a human-understandable explanation of the system's decision-making process. This can allow us to understand the basis for a network security alert and determine whether it is a genuine threat or a false positive. XAI can also be used to analyze the output of an IPS and identify patterns or trends that may indicate the presence of an intruder or other security threat.
Another way that XAI can assist IPSs is by helping to identify and prioritize potential threats. For example, an XAI system might be trained to recognize patterns of system usage that are typical for a particular user or group of users, and use this knowledge to identify deviations from these patterns that could indicate an intruder. XAI can also be used to analyze the impact of a potential threat and provide a risk assessment, which can help security analysts determine the appropriate response to the threat. In \cite{iadarola2021towards}, the authors suggest a method for Android malware detection and family identification that relies on application representation in terms of photos used to input an explainable deep learning model created by the authors. Furthermore, they demonstrate how the analyst may use explainability to evaluate alternative models. A data-driven IDS could incorrectly classify network attacks using the approach presented in \cite{marino2018adversarial}. A dynamic adversarial method is employed to find the minimum changes needed (of input features) to correctly classify a set of misclassified samples. It is through the magnitude of this modification that the most relevant features that explain the misclassification are identified. This methodology provides reasonable explanations describing the reasoning behind the misclassifications, and a description that matches expert knowledge.

\subsection{Access Control}
Access control is the act of restricting or denying access to a network to a user or to a group of users, typically for the purpose of preventing, limiting, or eliminating the ability of the user or the group to use, modify, or correct data or information. Access control is a legal concept that provides an objective measurement of the network security of a user's computer system, network, or communication system. Access control is often used in the context of protecting information or transactions in a computing system. 
Moreover, Access control-based XAI is the technical approach to ensuring that the right users have access to the right information at the right time. This involves access to data, applications, services and systems. 
Using one and two-dimensional CNNs, Adaptive Boosting, random forests, and KNN models, the authors in \cite{al2021explainable} propose a system to help manage access control and detect potential insider threats within an industrial internal security framework based on electrocardiogram and electroencephalogram (EEG) brainwave signals. According to this system, human attention states can be classified into four risk categories ranging from low to high risks. Within the same context of access control, a Focused Layer-wise Relevance Propagation (FLRP) approach is described in \cite{seibold2021focused}. With this framework, to secure facial images-based access control systems, a person can determine on a pixel level which image regions are used by the Deep Neural Network (DNN) to determine whether the face being examined is genuine or morphed. Also, the authors offer an additional framework for objectively analyzing the quality of their method and comparing FLRP to other interpretation methods based on DNN-based face morphing attack detectors. By removing detected artifacts, the evaluation framework aims to analyze how these changes affect the decisions of the DNN. 

\subsection{Authentication}
Authentication is one of the most fundamental aspects of the digital era. It is the process by which a person or organisation is recognised by another party for what they are and are not. It is used for the most basic level of authentication of information, which is to ensure the integrity of information. This has been used to provide a secure environment where information can be accessed, and made available. XAI can be used to assist with authentication by providing transparent and interpretable explanations for the decisions made by AI systems. This can be particularly useful in scenarios where the authentication process involves making decisions based on complex data or patterns that may be difficult for humans to understand or interpret.
For example, XAI can be used to help authenticate network users based on their behavior or characteristics. For example, an AI system might be trained to recognize patterns in a user's typing or mouse movement that are indicative of that user's identity. By providing explanations for the decisions made by the AI system, XAI can help users better understand how the authentication process works and increase their trust in the system.
In addition to assisting with user authentication, XAI can also be used to explain the AI decisions used to authenticate, in a more clear way, the authenticity of documents, images, or other types of data. For example, an AI system might be trained to detect forged or tampered documents by analyzing the patterns and features present in the document. By providing explanations for the decisions made by the AI system, XAI can help users better understand how the authentication functions and increase their confidence in the results.
Overall, XAI has the potential to improve the accuracy and reliability of authentication processes, as well as increase user trust and understanding of these systems.
In addition, many studies have pointed to authentication as a key factor in improving trust within the context of explainability and transparency.
As an example, the authors of the paper \cite{garcia2018explainable} demonstrate that fingerprinting and biometric authentication systems can be mimicked to produce unique signatures. Using XAI techniques, they construct a blind attack based on the query synthesis framework and expose the ineffectiveness of underlying ML classification models. By combining behavioral biometrics, ML, and domain-dependent aspects, the authors in \cite{rocha2021continuous} propose a continuous authentication approach that enables users to interpret the actions and decisions of the system. This method is non-intrusive, does not require any additional hardware, and can be used continuously to monitor user identity.

\begin{table*}[!htp]
\resizebox{0.75\textwidth}{!}
{
\begin{minipage}
{\textwidth}
\begin{center}
\caption{Comparative summary of the main XAI techniques for  cybersecurity}
\label{table777}
\hspace*{-77.0pt}
\begin{tabular}{c c c c }
\hline\hline
Approach & XAI Method Level & Cybersecurity Class & Technique  \\ 
\hline
Harder et al. \cite{harder2020interpretable} &Local explanation& Privacy&locally linear maps (LLM)  \\
L{\"o}bner et al. \cite{lobner2021explainable} & post-hoc explanation, Model-agnostic explainer &Privacy&
\shortstack{Information Gain (IG) and \\  extended Iterative Dichotomiser 3 (ID3) classification tree}\\
Rjoub et al. \cite{rjoub2020trust}&Local explanation, Post-hoc explanation &Trust&Double Deep Q Learning (DDQN) \\
Rjoub et al. \cite{rjoub2022explainable}  & Local explanation, Post-hoc explanation & Trust& SHAP  \\
Machlev et al. \cite{machlev2021measuring}  & Model-agnostic explainer, Post-hoc & Trust &Grad-CAM, LIME\\
Kuppa et al. \cite{kuppa2020black}  & post-hoc explanation, Model-agnostic explainer & Trust &   \shortstack{Input*Gradient(I*G), Layer-Wise Relevance Propagation(LRP), \\ Guided Back Propagation(GBP),\\ Smooth- Grad(SG), Gradient(GRAD), and Integrated Gradients(IG)}\\
Mankodiya et al. \cite{mankodiya2021xai}  & Model-specific explainers & Trust & Decision Tree-based algorithms\\
Elayan et al. \cite{elayan2021internet}  & post-hoc explanation & Trust&IoB-XAI \\
R. Karn et al. \cite{karn2020cryptomining}  & Local explanation, post-hoc explanation & ID &SHAP \& LIME \\
M. Wang et al. \cite{wang2020explainable}  & Local \& Global explanation & ID &SHAP \\
G. Baryannis et al. \cite{baryannis2019predicting}  & Model-specific & ID & Decision Tree \& SVM \\
L. Aguilar. \cite{aguilar2022towards}  & Local explanation, post-hoc explanation & ID & Decision Tree \\
G. Iadarola et al. \cite{iadarola2021towards}  & Global explanation, Model-specific & Intrusion prevention & Gradient-weighted Class Activation Mapping (Grad-CAM) \\
L. Marino et al. \cite{marino2018adversarial}  & Local explanation, Model-specific, post-hoc explanation & Intrusion prevention & Adversarial machine learning \\
Y. Al Hammadi et al. \cite{al2021explainable}  & Local explanation, pre-hoc explanation,  Model-specific & Access Control & SHAP \\
C. Seibold et al. \cite{seibold2021focused}  & Global explanation, Post-hoc explanation & Access Control & Focused Layer-wise Relevance Propagation (FLRP) \\
W. Garcia et al. \cite{garcia2018explainable}  & Model-agnostic explainer, Post-hoc explanation, Local explanation & Authentication & LIME \\
R. Rocha et al. \cite{rocha2021continuous}  & Global explanation, Model-agnostic explainer, Post-hoc explanation,  & Authentication & Gedeon method \\
\hline
\end{tabular}
\end{center}

\end{minipage} 
}
\end{table*}

Table \ref{table777} presents a comparative summary of XAI techniques used in cybersecurity, along with the approach, XAI method level, and cybersecurity class. The table includes various studies that have used different XAI techniques, such as local explanation, global explanation, post-hoc explanation, pre-hoc explanation, and model-specific explainers to analyze  cybersecurity issues according to our classification introduced in Section \ref{Papers_Classification}, which includes  privacy, trust, intrusion detection and prevention, access control, and authentication.

\section{Desirable Criteria for Future Solutions}
\label{criteria}
Our classification and discussion above have led us to identify key crucial criteria that, in our opinion, ought to be taken into account when developing future solutions for XAI and  cybersecurity. We will present and debate how these criteria may be practically applied to design effective solutions in the areas of XAI and cybersecurity, using these criteria as our guide. 

\begin{itemize}
    \item \textbf{Criterion \#1}: \textit{Transparency}: the ability of human agents to gain insight into the inner workings of an XAI, and decision-making processes of an AI system. XAI systems must be open and available to all concerned parties and must clearly disclose their capabilities, limitations, potential impacts, modes of operation, data sources, and collection mechanisms for underlying data or information. As an example in IDS, IPS, etc., lack of explainability limits the security analyst’s ability to understand why the system has generated a certain output. Explainability refers to the ability of an AI system to provide human-readable explanations for its decisions, while predictive performance refers to the accuracy of an AI system's predictions or recommendations. On one hand, explainability is important because it helps us build trust and accountability in the use of AI, and can assist in ensuring that AI systems are being used ethically and responsibly. On the other hand, predictive performance is important because it determines the usefulness and effectiveness of an AI system. In some cases, it may be possible to trade off some level of predictive performance in order to increase explainability, but in other cases, there may be a significant trade-off between the two. For example, some AI algorithms are highly complex and difficult to interpret, but they may also have very high predictive performance. In these cases, it may be difficult to increase the explainability of the AI system without significantly reducing its predictive performance.

    In many situations, there is a tension between explainability and predictive performance: models that are more complex and opaque are often more accurate in their predictions than models that are simpler and more transparent.  It is important to understand that explainability is not an all-or-nothing property: there are many degrees of explainability, and a model can be more or less explainable.

    \item \textbf{Criterion \#2}: \textit{Understandability}: a domain cybersecurity expert must be able to understand the outputs and results of an XAI system. Trust, for example, is based on many heuristics, so a domain expert might want to understand how an XAI system is making its decisions and providing its explanations. If a domain expert does not understand how an XAI system arrives at its results, then a lack of trust in the system will be established. Additionally, explainability also allows for modifications to the XAI system. If a domain expert does not understand why the system is misclassifying inputs, they will not be able to modify the system to correct these errors.

    \item \textbf{Criterion \#3}: \textit{Cogency}: the XAI system must make sense, and be able to explain itself so that even non-experts can understand its behavior and predictions. For example in intrusion detection, explainability is important for demonstrating why a certain event is benign or malicious. Without explainability, the system could flag benign behavior as malicious and block a user from accessing a system. The system should also be able to provide an explanation of why the benign behavior was flagged as malicious, and why the malicious behavior was not flagged. If the system cannot provide cogent explanations, it will not be trusted by users, and might not be used.

    \item \textbf{Criterion \#4}: \textit{Broadness}: XAI system must handle many different cybersecurity cases and be applicable to a wide variety of domains, to be truly useful as a general intelligence tool. For example, in privacy, the system should be flexible enough to work with a wide range of data sets, data types, and domains. Additionally, it must be able to cope with rapidly evolving threats.

    \item \textbf{Criterion \#5}: \textit{Autonomy}: XAI systems must be able to act autonomously (behave in their own way) without human supervision. For example in an access control system,  if the system is not autonomous, when an attacker tricks the system into letting them in, then a human needs to be there to catch the mistake. However, there are also potential risks associated with autonomy in AI. For example, if an AI system is not designed or implemented properly, it may make decisions that are inappropriate or even harmful. This is especially true if the AI system is being used in a safety-critical application, such as an access control system. To address these risks, it is important to ensure that AI systems are designed and implemented in a way that is transparent and explainable (XAI). This is highly useful to build trust and accountability in the use of the AI system, and can help us ensure that the AI system is acting in a safe and responsible manner. XAI systems need to be able to detect and correct possible mistakes while explaining the decisions made on their own. 
    \item \textbf{Criterion \#6}: \textit{Openness}: XAI systems must provide sufficient access to their methods, predictions, and decision making processes so that they can be properly evaluated, improved, and checked for consistency. For example, XAI systems may be used to analyze large amounts of data to detect malicious intent or behavior. If users of such systems cannot verify that the system is detecting malicious behavior effectively, then they will have little confidence in the system and it will be much less useful. Users must have access to the system’s data analysis methods so that they can examine the system’s performance themselves and decide whether or not to trust it. Additionally, if a system is found to have biases, users must be able to easily modify the system to remove them.
    \item \textbf{Criterion \#7}: \textit{Reasonableness}: XAI systems must show concern for the common good and respect human life, systems, and the surrounding environment avoiding prejudicial actions or behaviors that would harm or destroy shared resources and private data. For example, cybersecurity concerns, such as viruses and malware, must be carefully managed to ensure that they do not cause unnecessary harm or destruction. This includes ensuring that these types of threats are not indiscriminate, and are targeted specifically at addressing a specific security concern. Furthermore, data collection that is not based on a reasonable legal basis must be carefully managed to ensure that it is proportionate to the aim being pursued and does not infringe on the privacy of individuals. This includes taking steps to protect the privacy and security of personal data, and being transparent about how the data is being collected and used. Overall, it is highly significant to ensure that AI systems are designed and implemented in a way that is responsible, transparent, and respectful of the privacy and security of individuals. This includes taking steps to address cybersecurity concerns and managing data collection in a way that is ethical and compliant with relevant laws and regulations. However, there are a number of exceptions to the rule, such as when the intrusion or unauthorized access threatens public safety or national security.
    \item \textbf{Criterion \#8}: \textit{Coping}: XAI systems must be capable of learning and adapting to dynamic, complex, and rapidly changing environments in order to provide effective solutions, especially if they are going to perform essential roles as servants and caregivers in real-world environment (eg. health care). For example, the environment can rapidly change and be evolving at an alarming rate. In IDSs, there is need to rapidly adapt to new threats and be able to protect against them. In particular, XAI systems must rapidly adapt to changes and be capable of learning from them to provide effective explainable solutions. However, this is not an easy task. The systems need to be constantly learning and be given the ability to update their knowledge. 
    \item \textbf{Criterion \#9}: \textit{Responsibility}: XAI systems must be capable of being held accountable for their actions and, where appropriate, to take action to ensure the safety of humans, systems, and the implementation environment. For example, the system should regularly assess the risk posed by users and content and take appropriate action, such as blocking access to potentially harmful content, suspending or terminating users’ accounts, or reporting the issue to law enforcement.
\end{itemize}

 The current literature on cybersecurity concerns in XAI can be improved in many different respects. To begin with, the ability of human agents to gain insight into the inner workings of an XAI should be made available by instructing the system to provide information about its functioning. More elaborate developments are needed with respect to the different kinds of information that can be imparted by an XAI in order to adequately facilitate cyber risk assessment, management, and mitigation. Several XAI methodologies that have been utilized by several approaches \cite{al2022xai,hariharan2022xai} in the field of cybersecurity provide a significant potential in the direction of achieving this goal. However, further research on the interdependencies between this technology and the other components of XAI, such as the cost and the assurances of security and data privacy, is required. A further aspect of research involves gaining a deeper understanding of the process an XAI system goes through in order to create an output. XAI output and results are two aspects that need to be considered in order to determine if a technology fits its intended use. In fact, the majority of these approaches, despite the importance of looking over to understand the outputs and results of an XAI system, overlook the cybersecurity issues. This is the case despite the fact that understanding the outputs and results of a XAI system is important. However, the vast majority of contemporary deep learning techniques, such as deep neural networks, which serve as the paradigm's central support structure, are founded on black-box models.
Another aspect to consider in the future is exploring the ability of the XAI system to explain itself so that even non-experts can understand its behavior and predictions. As cybersecurity experts and non-experts work with this type of system more, there will be a desire to understand what XAI is doing, and why it is making the decisions that it does, and how these decisions are affecting the system.

 The XAI's ability to handle different cybersecurity cases and be applicable to a wide variety of domains is another important aspect that needs careful investigation in the future. To achieve that ability, the future XAI-based cybersecurity models need to extend its architecture with a few more capabilities such as data mining, pattern recognition, and intelligent systems. Moreover, it is required for an intelligent cybersecurity system to understand its environment and learn from it before being able to make predictions about future events or transfer the knowledge to another systems and environments. XAI systems must be able to act autonomously without human supervision in order to perform tasks effectively, to deal with changing environments or conditions, and to learn on their own by interacting with their environments. Thus, a major challenge for XAI systems is designing how to learn from this interaction and store or remember those experiences. Another research direction would be to provide sufficient access to XAI methods, and their predictions, and decision making processes, so that they can be properly evaluated, improved, and checked for consistency. This requires considerable work by the researchers that have developed and use XAI methods in cybersecurity field. In addition, since there is an absolute dearth of research on the evaluation of XAI-based cybersecurity methods, this may be the right time to move ahead with XAI research. 

The XAI system must also be developed in a way that is reasonable, balancing the need to protect human life and privacy while ensuring the system is effective and efficient. XAI systems that are designed without these considerations in mind may cause harm to individuals or society.  For example, an autonomous vehicle that is not programmed to avoid harming pedestrians may cause injuries or fatalities if it collides with a pedestrian.  A facial recognition system that does not take into account the rights of individuals to privacy may violate those rights if it is used to collect and store biometric data without the individual’s consent. On the other hand, as XAI systems increasingly become a part of our lives and daily routines, it is important for them to be able to coping with dynamic, complex, and rapidly changing environments. In order for XAI systems to be able to cope with dynamic and complex environments, they must be able to learn and adapt. This is especially important if they are going to serve as essential caregivers in a real-world environment. Moreover, in order to ensure the safety of humans, systems and the implementation environment, XAI systems must be capable of being held accountable for their actions and, where appropriate. Table \ref{table:criteria} summarizes the main approaches to XAI challenges in cybersecurity and highlights the criteria that each underlying approach meets. 

\begin{figure*}[!ht]
 \includegraphics[width=\textwidth,height=.99\textwidth]{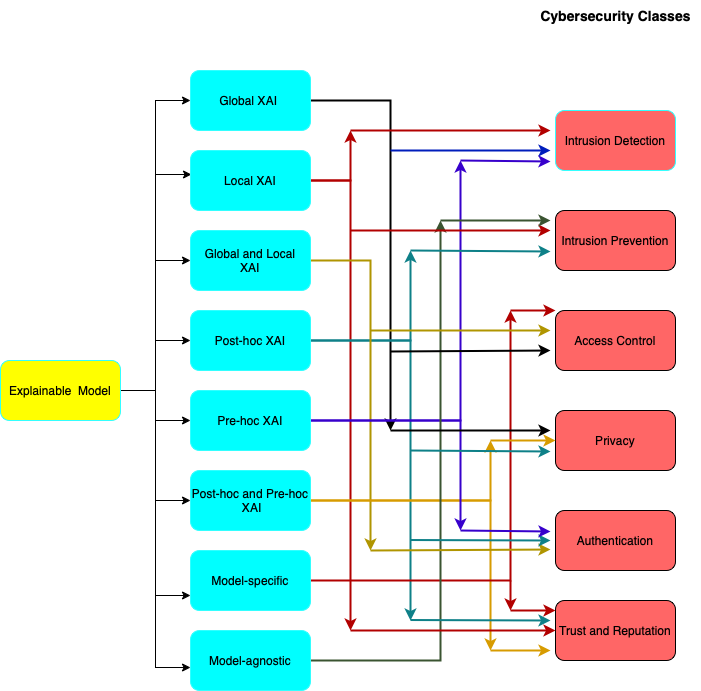}
  \caption{Linking XAI models to cybersecurity classes}
  \label{fig797}
\end{figure*}

\begin{figure*}[!ht]
 \includegraphics[width=\textwidth,height=.99\textwidth]{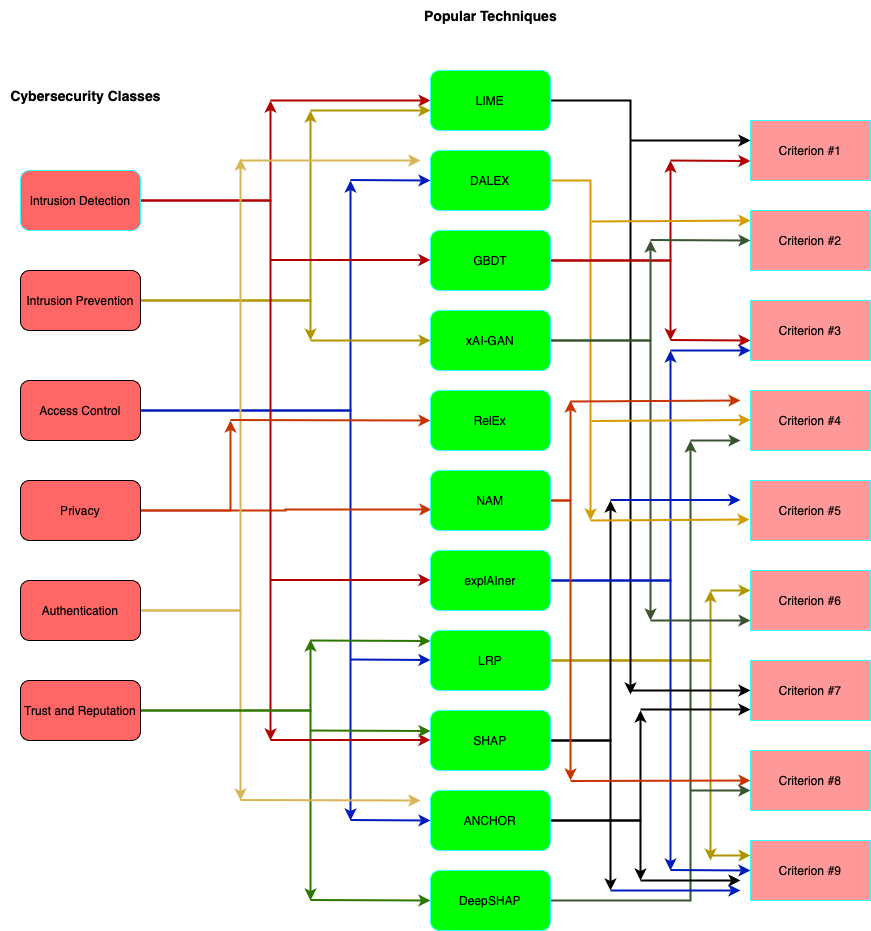}
  \caption{Linking cybersecurity classes to popular XAI techniques and desirable criteria for future solutions}
  \label{fig799}
\end{figure*}

In addition, Figs. \ref{fig797} and \ref{fig799} provide detailed recommendations for future research directions in the field of XAI for cybersecurity. 
In particular, Fig. \ref{fig797} illustrates how XAI models are linked to cybersecurity classes. It provides a visual representation of the relationships between different XAI models and their applications to various cybersecurity classes such as privacy, trust, IDS, intrusion prevention, authentication, and access control. The figure can serve as a guide for researchers and practitioners to understand the relevant XAI techniques that can be used for specific cybersecurity classes. On the other hand, Fig. \ref{fig799} shows how cybersecurity classes are mapped to popular XAI techniques and desirable criteria for future solutions. The figure provides a comprehensive view of the desirable characteristics for XAI solutions in the field of cybersecurity. It presents a mapping of different cybersecurity classes to popular XAI techniques, such as SHAP and LIME, and also identifies desirable criteria for future solutions such as Criterion $\#2$ (Understandability), Criterion $\#7$ (Reasonableness), and Criterion $\#9$ (Responsibility). This figure can be used as a reference to identify the appropriate XAI techniques and desirable criteria when developing XAI solutions for cybersecurity.


\begin{table*}[!htp]
\resizebox{0.65\textwidth}{!}
{\begin{minipage}
{\textwidth}
\caption{ Summary of the main XAI techniques and criteria for  cybersecurity}
\hspace{-2.5cm}
\small
{\large %
\begin{tabular}{c c c c c }

\hline\hline
Approach &  Cybersecurity Class & Technique & Main Idea & Criteria  \\ 
\hline
\hline
\small{Harder et al. \cite{harder2020interpretable}} & \small{Privacy} & \small{Locally Linear Maps (LLM)} &\begin{tabular}{l}
\begin{minipage}[t]{0.5\columnwidth}%
\small{To achieve high classification accuracy and differentially private explanations, the authors construct a family of simple models with the aim of approximating complex models by using several locally linear maps per class.}  
\end{minipage}\tabularnewline
\end{tabular}  & \small{Criteria \#2} \\
\small{L{\"o}bner et al. \cite{lobner2021explainable}} & \small{Privacy} &\small{ \shortstack{Information Gain (IG) and \\ extended Iterative Dichotomiser 3 (ID3) classification tree}} &\begin{tabular}{l}
\begin{minipage}[t]{0.5\columnwidth} \small{For default privacy setting prediction, the authors provide enhanced feature selection, improved interpretability, and enhanced evaluation metrics that are easy to use and therefore easier to understand.}  \end{minipage}\tabularnewline
\end{tabular} & \small{Criteria \#2, and \#8} \\
\small{Rjoub et al. \cite{rjoub2020trust}}& \small{Trust} &  \small{Double Deep Q Learning (DDQN)} &  \begin{tabular}{l}
\begin{minipage}[t]{0.5\columnwidth} \small{The authors argue that trust should be an integral part of the decision-making process and therefore design a trust establishment mechanism between the edge server and IoT devices.} \end{minipage}\tabularnewline
\end{tabular} & \small{Criteria \#2, and \#8} \\
\small{Rjoub et al. \cite{rjoub2022trust}}  & \small{Trust} & \small{SHAP} & \begin{tabular}{l}
\begin{minipage}[t]{0.5\columnwidth} \small{The authors design an Explainable ArtificialIntelligence (XAI) Federated Deep Reinforcement Learningmodel to improve the effectiveness and trustworthiness of thetrajectory decisions for newcomer Autonomous Vehicles (AVs).} \end{minipage}\tabularnewline
\end{tabular} & \small{Criteria \#1, and \#5} \\
\small{Machlev et al. \cite{machlev2021measuring} } & \small{Trust} & \small{Grad-CAM, LIME} & \begin{tabular}{l}
\begin{minipage}[t]{0.5\columnwidth} \small{ Explainable artificial intelligence is used to explain PQD classifier results (XAI). XAI approaches and classifiers are merged and graded based on their explanations during validation.}  \end{minipage}\tabularnewline
\end{tabular} & \small{Criteria \#9} \\
\small{Kuppa et al. \cite{kuppa2020black}}  & \small{Trust} & \small{\shortstack{Input*Gradient(I*G), Layer-Wise Relevance Propagation(LRP), \\ Guided Back Propagation(GBP),\\ Smooth- Grad(SG), Gradient(GRAD), and Integrated Gradients(IG)}} & \begin{tabular}{l}
\begin{minipage}[t]{0.5\columnwidth}\small{The authors propose a taxonomy for Explainable Artificial Intelligence (XAI) by designing a novel black-box attack based on gradient based XAI methods to study their consistency, correctness, and confidence properties.} \end{minipage}\tabularnewline
\end{tabular} & \small{Criteria \#3, and \#6} \\
\small{Mankodiya et al. \cite{mankodiya2021xai}}  & \small{Trust} & \small{Decision Tree-based algorithms} & \begin{tabular}{l} \begin{minipage}[t]{0.5\columnwidth} \small{The authors propose a machine learning approach to detect misbehaving vehicles in a VANET. To classify the attacker AV, they used decision tree-based algorithms.} \end{minipage}\tabularnewline
\end{tabular} & \small{Criteria \#7} \\
\small{Elayan et al. \cite{elayan2021internet} } & \small{Trust} & \small{IoB-XAI} &  \begin{tabular}{l} \begin{minipage}[t]{0.5\columnwidth} \small{For influencing IoT behavior, the authors propose the internet of behavior (IoB) and explainable AI systems.} \end{minipage}\tabularnewline
\end{tabular}  & \small{Criteria \#4, and \#9} \\
\small{R. Karn et al. \cite{karn2020cryptomining}}  & \small{ID} & \small{SHAP \& LIME} &  \begin{tabular}{l} \begin{minipage}[t]{0.5\columnwidth} \small{In this paper, ML-based explainable models are used to detect cryptomining applications in a Kubernetes cluster.} \end{minipage}\tabularnewline
\end{tabular}  & \small{Criteria \#9} \\
\small{M. Wang et al. \cite{wang2020explainable}}  & \small{ID} & \small{SHAP} & \begin{tabular}{l} \begin{minipage}[t]{0.5\columnwidth} \small{To improve understanding of IDSs, the authors propose combining local and global explanations with SHapley Additive Explanations (SHAP).} \end{minipage}\tabularnewline
\end{tabular} & \small{Criteria \#2} \\
\small{G. Baryannis et al. \cite{baryannis2019predicting}}  & \small{ID} & \small{Decision Tree \& SVM}
& \begin{tabular}{l} \begin{minipage}[t]{0.5\columnwidth} \small{Researchers propose a framework for SCRM risk prediction based on data-driven AI techniques, which collaborates and interacts with supply chain experts.} \end{minipage}\tabularnewline
\end{tabular} & \small{Criteria \#3} \\
\small{L. Aguilar. \cite{aguilar2022towards} } & \small{ID} & \small{Decision Tree} & \begin{tabular}{l} \begin{minipage}[t]{0.5\columnwidth} \small{Using decision trees, the authors propose an interpretable autoencoder, which does not require data transformation when handling categorical data.} \end{minipage}\tabularnewline
\end{tabular}  & \small{Criteria \#9} \\
\small{G. Iadarola et al. \cite{iadarola2021towards}}  & \small{Intrusion prevention} & \small{Gradient-weighted Class Activation Mapping (Grad-CAM)}  & \begin{tabular}{l} \begin{minipage}[t]{0.5\columnwidth} \small{For Android malware detection, the authors propose an explanation-based deep learning model based on application representation in terms of images.} \end{minipage}\tabularnewline
\end{tabular}  & \small{Criteria \#4, and \#7} \\
\small{L. Marino et al. \cite{marino2018adversarial}}  & \small{Intrusion prevention} & \small{Adversarial machine learning} & \begin{tabular}{l} \begin{minipage}[t]{0.5\columnwidth} \small{Researchers present an approach to explain incorrect classifications made by data-driven intrusion detection systems (IDSs).} \end{minipage}\tabularnewline
\end{tabular}  & \small{Criteria \#3} \\
\small{Y. Al Hammadi et al. \cite{al2021explainable}}  & \small{Access Control} & \small{SHAP} &  \begin{tabular}{l} \begin{minipage}[t]{0.5\columnwidth} \small{The authors present an insider risk assessment system as a fitness for duty security evaluation using EEG brainwave signals with explainable deep learning and machine learning algorithms for classifying abnormal EEG signals which indicate a potential insider threat and evaluating fitness for duty.} \end{minipage}\tabularnewline
\end{tabular} & \small{Criteria \#8, and \#9} \\
\small{C. Seibold et al. \cite{seibold2021focused}}  & \small{Access Control} & \small{Focused Layer-wise Relevance Propagation (FLRP)} & \begin{tabular}{l} \begin{minipage}[t]{0.5\columnwidth} \small{Using this framework, the authors explain to a human inspector at a pixel level which regions are used by a DNN to distinguish between a genuine and a morphed face image.} \end{minipage}\tabularnewline
\end{tabular}  & \small{Criteria \#3} \\
\small{W. Garcia et al. \cite{garcia2018explainable}}  & \small{Authentication} & \small{LIME} & \begin{tabular}{l} \begin{minipage}[t]{0.5\columnwidth} \small{Researchers explore how adversaries can infer decision boundaries from victim models by using the XAI approach.} \end{minipage}\tabularnewline
\end{tabular} & \small{Criteria \#1, and \#2} \\

\small{R. Rocha et al. \cite{rocha2021continuous}}  & \small{Authentication} & \small{Gedeon method} & \begin{tabular}{l} \begin{minipage}[t]{0.5\columnwidth} \small{Based on behavioral biometrics and machine learning, the authors propose an approach for continuous authentication that includes domain-dependent aspects for the user to interpret the actions and decisions of the system.} \end{minipage}\tabularnewline
\end{tabular} & \small{Criteria \#5} \\\\
\hline
\end{tabular}
}
\label{table:criteria}
\end{minipage} }
\end{table*}

Additionally, the criteria presented in this section should not be considered as an exhaustive list of all possible desirable properties of XAI solutions in the context of cybersecurity. As XAI technology continues to evolve and new security threats emerge, new criteria may need to be identified and incorporated into the evaluation process. Therefore, it is important for researchers and practitioners in this field to remain vigilant and adapt to changes in the threat landscape to ensure the continued effectiveness and relevance of XAI solutions for cybersecurity. It is also crucial to promote collaboration and knowledge-sharing between stakeholders, including academia, industry, and government agencies, to facilitate the development and deployment of XAI solutions that meet the evolving needs of the cybersecurity community. By continuously reviewing and updating the criteria for evaluating XAI solutions in cybersecurity, we can ensure that these solutions are effective, efficient, and trustworthy in addressing emerging threats and challenges in this critical area.

\section{Future Directions}
\label{future}
In this paper, we have surveyed key concerns for effective
XAI in the domain of cybersecurity applications.
We have also provided important characterizations of the central methods to be considered and evaluated.
In this section, we step back to discuss what we feel are
some of the most critical future research paths to pursue,
in order for efforts on explainable AI for cybersecurity to achieve success.
 
\subsection{Risk of Machine Interpretation in Cybersecurity}

The goal of researching machine interpretation in the context of  cybersecurity is to improve the reliability and accuracy of AI systems when interpreting and analyzing data related to cybersecurity threats. This can involve developing and evaluating new algorithms, techniques, and approaches for interpreting data from a variety of sources, including network traffic and logs, and user behavior data.
One key challenge in this area is to develop machine interpretation techniques that are able to accurately identify and classify  cybersecurity threats, while also minimizing the risk of false positives (incorrectly identifying a threat that does not actually exist). This can involve developing techniques for analyzing the context and background information related to a given data source, as well as using machine learning algorithms to identify patterns and trends that may indicate the presence of a cybersecurity threat.

 In particular, we would like to answer the following questions: (1) What are the limitations of current machine interpretation methods in the cybersecurity domain? (2) How can we design more effective machine interpretation methods that are better suited for the cybersecurity domain? (3) How can we improve the usability of machine interpretation methods in the cybersecurity domain? (4) What are the challenges of deploying machine interpretation methods in the cybersecurity domain? In order to answer these questions, we need to better understand the strengths and limitations of different machine interpretation methods. In addition, we need to develop new evaluation metrics that are better suited for measuring the performance of machine interpretation methods in the cybersecurity domain. Finally, we need to investigate how to improve the usability of machine interpretation methods in the cybersecurity domain.  

\subsection{Transparency of XAI Methods in Cybersecurity}
Given the importance of transparency in the field of  cybersecurity, there is a clear need for XAI methods that can provide understandable explanations of their decision-making processes. However, designing XAI methods that are both effective and transparent is a non-trivial task. In particular, the trade-off between accuracy and transparency is a major challenge that must be addressed. One potential solution is to design XAI methods that can provide different levels of transparency. For example, an XAI system could provide a high level of transparency when it is used to make low-risk decisions, and a low level of transparency when it is used to make high-risk decisions. This would allow users to trade off between accuracy and transparency, depending on their needs. Another potential solution is to use transparency-enhancing techniques, such as sensitivity analysis, to improve the transparency of XAI methods without sacrificing accuracy. In general, the development of XAI methods that are both effective and transparent is a major challenge that must be addressed in order to ensure the widespread adoption of XAI in the cybersecurity domain.

\subsection{Uncertainty Handling in XAI for Cybersecurity}

Given the highly uncertain and dynamic nature of the cybersecurity domain, there is a clear need for XAI methods that can deal with uncertainty. One potential solution is to use probabilistic models, such as Bayesian networks, to represent and reason about uncertainty. Another potential solution is to use robust optimization techniques to handle uncertainty in a principled way. In general, the development of XAI methods that can deal with uncertainty is a major challenge that must be addressed in order to ensure the widespread adoption of XAI in the cybersecurity domain.

\subsection{Secure and Efficient Algorithms for Distributed XAI Training}

When training XAI models, it is often necessary to use a large amount of data so that more useful explanations can be provided. This can be a problem in the network and system security domains, where data is often sensitive and distributed across different networks and organizations. A potential solution is to use federated learning, which is a distributed ML technique that allows multiple organizations to train a shared model without sharing their data. Federated learning has several advantages, including improved security and privacy, and reduced training time. However, federated explainable learning can also be a challenge, due to the need to design explainable algorithms that are secure and efficient. In general, the development of efficient and secure algorithms for distributed XAI training is a major challenge that must be addressed in order to ensure the widespread adoption of XAI in the cybersecurity domain. On the other hand, designing explainable federated learning models is more challenging than traditional deep learning models because federated learning involves training multiple local models on different data sources, and then aggregating these models to produce a global explainable model. This distributed nature of federated learning can make it more difficult to understand and explain the decision-making process of the global model.
To address this challenge, it is necessary to ensure that both the local models and the aggregation process are explainable. This can involve using techniques such as feature importance analysis or saliency maps to understand how the local models are making decisions, and developing techniques for aggregating the local models in a way that is transparent and easy to understand.

\subsection{Cybersecurity-Specific Evaluation Metrics}
Developing proper evaluation metrics is critical for the success of machine interpretation methods in any domain. However, it is especially important in the cybersecurity domain due to the unique nature of the data and tasks involved. For instance, many existing evaluation metrics (e.g., accuracy) are not well suited for measuring the performance of machine interpretation methods on cybersecurity data due to its unbalanced class distribution and high dimensional space (i.e., multiple features/variables). In addition, most existing evaluation metrics do not take into account how different types of errors can have different consequences in the cybersecurity domain. For example, a false positive error (i.e., an alert that incorrectly identifies a malicious activity) might cause unnecessary alarm while a false negative error (i.e., failing to detect a real attack) can result in serious damage or even loss of life.

Certainly, in addition to the future directions discussed above, there are several other areas of research that could be explored in the context of XAI and cybersecurity. For example, there is a need for further investigation into how XAI techniques can be applied to the detection of highly sophisticated adversarial attacks and advanced persistent threats (APTs), which are often challenging to identify using traditional security tools. Engineered adversarial attacks and APTs are persistent, targeted attacks that are typically carried out by skilled attackers over an extended period of time, often using a range of sophisticated tactics to evade detection.
Moreover, there is a growing need to develop XAI techniques that can handle large-scale, real-time data streams, as many  cybersecurity applications involve analyzing vast amounts of data in real-time. Additionally, there is a need to explore how XAI techniques can be used to improve the usability and effectiveness of cybersecurity tools, such as intrusion detection systems, by providing users with more meaningful and actionable insights into security incidents.
Another area of research that could be explored is the development of XAI techniques that can operate in decentralized and distributed environments, such as those found in many Internet of Things (IoT) applications. In such environments, data is often generated and processed across a large number of heterogeneous devices, which presents significant challenges for traditional cybersecurity approaches. XAI techniques that can operate in these environments could provide significant benefits in terms of improving the accuracy and timeliness of threat detection and response.
Finally, there is a need for further research into the ethical and societal implications of XAI in cybersecurity. As XAI techniques become more prevalent in the field, it is essential to consider the potential risks and unintended consequences of their use, such as the potential for XAI systems to be biased or to reinforce existing power imbalances. It is also essential to ensure that XAI systems are transparent, explainable, and accountable, to enable effective human oversight and ensure that their use aligns with ethical and legal norms.

\section{CONCLUSION}
\label{conc}
In this paper, we have, first, discussed the motivation for approaches that aim to achieve XAI, enabling users to trust and understand the AI systems that are in use today in many organizations.
We have also clarified specific challenges that arise for XAI when the AI systems in question are being used for cybersecurity. In order to ground our survey of various models that are in use, we then distinguished a core collection of key terms that are of concern. Following this, we made clear how our particular survey differs from other efforts to date that summarize XAI approaches in the literature. We then situated ourselves squarely in our area of focus, XAI for cybersecurity, outlining some well-founded decisions for what to study, how to characterize the field and and how best to present our results.

Considerable depth is introduced with a clarification of the ways in which mathematical structure can be considered, as well as what kinds of classification of methods are most effective in order to compare approaches.
We reflected at great length on the most important considerations for the cybersecurity application, leading to a comparative summary of the main XAI techniques for systems that are designed for this domain.
A key contribution is our specific analysis that proposes the primary
criteria to resolve, in order to achieve successful XAI  cybersecurity solutions. This is followed by a deep discussion of key considerations for future work, to achieve these goals.

In all, we expose the community of AI researchers to a specific
area of great value for the future: continuing with current efforts
to engender trust of AI systems through XAI, but with a new awareness of the landscape of methods and challenges, when  cybersecurity is the central concern of these AI systems.

\bibliographystyle{IEEEtran}
\bibliography{mybibliography.bib}

\begin{IEEEbiography}
[{\includegraphics[width=1in,height=1.25in,clip,keepaspectratio]{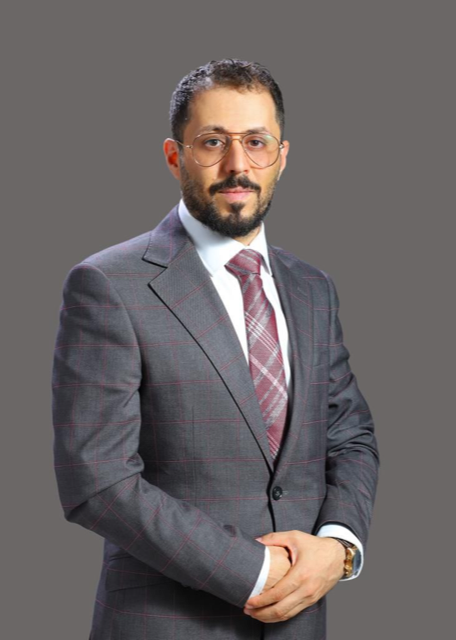}}]%
{Gaith Rjoub}
received his Master's degree in Quality Systems Engineering from Concordia Institute for Information Systems Engineering (CIISE), Canada in 2014, and his PhD degree in Information and Systems Engineering from CIISE, Canada in 2021. He is an assistant professor in the Data Science department at Princess Sumaya University for Technology (PSUT). From 2021 to 2022, he was a Postdoctoral Fellow with University of Montreal and Concordia University. His research interests include cloud and edge computing, machine and deep learning, artificial intelligence and big data analytics.
\end{IEEEbiography}

\begin{IEEEbiography}
[{\includegraphics[width=1in,height=1.25in,clip,keepaspectratio]{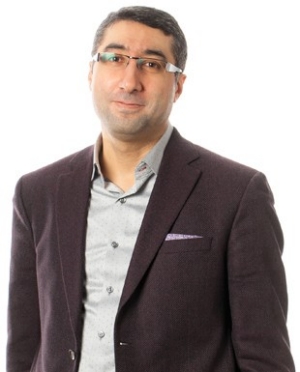}}]%
{Jamal Bentahar}  (Member, IEEE) received the Ph.D. degree in computer science and software engineering from Laval University, Québec, QC, Canada, in 2005.,He is a Professor with Concordia Institute for Information Systems Engineering, Concordia University, Montreal, QC, Canada. From 2005 to 2006, he was a Postdoctoral Fellow with Laval University, and then an NSERC Postdoctoral Fellow with Simon Fraser University, Burnaby, BC, Canada. He is a Visiting Professor with Khalifa University, Abu Dhabi, UAE. His research interests include the areas of computational logics, reinforcement learning, multiagent systems, service computing, game theory, and software engineering.,Prof. Bentahar was an NSERC Co-Chair for Discovery Grant for Computer Science from 2016 to 2018. He is an Associate Editor of IEEE TSC.
\end{IEEEbiography}

\begin{IEEEbiography}
[{\includegraphics[width=1in,height=1.25in,clip,keepaspectratio]{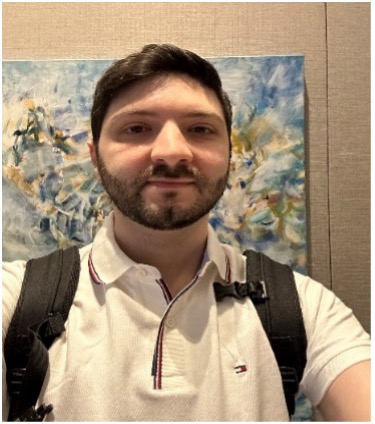}}]%
{Omar Abdel Wahab}
is Assistant Professor in the Department of Computer and Software Engineering at Polytechnique Montréal, Canada. He holds a
Ph.D. in Information and Systems Engineering from Concordia University,
Montréal, Canada. He received his M.Sc. in computer science in 2013 from
the Lebanese American University (LAU), Lebanon. From 2017 to 2018, he
was postdoctoral fellow at the École de Technologie Supérieure (Canada),
where he worked on an industrial research project in collaboration with
Rogers and Ericsson. His research interests are in the areas of cybersecurity,
artificial intelligence, Internet of Things and big data analytics.
\end{IEEEbiography}

\begin{IEEEbiography}
[{\includegraphics[width=1in,height=1.25in,clip,keepaspectratio]{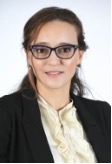}}]%
{Rabeb Mizouni} is an Associate Professor in the Department of Electrical Engineering and Computer Science at Khalifa University, Abu Dhabi, United Arab Emirates. She got her M.Sc. and Ph.D. in Electrical and Computer Engineering from Concordia University, Montreal, Canada in 2002 and 2007 respectively. Currently, she is interested in the deployment of context aware mobile applications, crowd sensing, Artificial Intelligence, IoT and Blockchain. Dr. Mizouni is currently an Associate Editor for IEEE Internet of Things magazine and the IEEE Transaction on Service Computing.
\end{IEEEbiography}

\begin{IEEEbiography}
[{\includegraphics[width=1in,height=1.25in,clip,keepaspectratio]{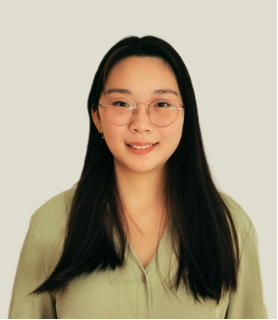}}]%
{Alyssa Song}  is an undergraduate student studying Computer Science at the University of Waterloo in Waterloo, Ontario Canada. She completed an undergraduate research assistantship
in Spring 2022 under the co-supervision of Professor Robin
Cohen and Professor Jamal Bentahar.
She was also a student that term in the Responsible AI program run under the Natural Sciences and Engineering Council of Canada's CREATE (Collaborative Research and Training Experience) initiative.
\end{IEEEbiography}

\begin{IEEEbiography}
[{\includegraphics[width=1in,height=1.25in,clip,keepaspectratio]{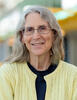}}]%
{ Robin Cohen} is a Professor in the Cheriton School of Computer
Science at the University of Waterloo in Waterloo, Ontario Canada.
She conducts research in the subfield of Artificial Intelligence
known as Multiagent Systems and has a particular interest in
exploring multiagent trust modeling, with applications to social media.
She is a Senior Member of the AAAI and a recipient of the Canadian
Artificial Intelligence Association's Lifetime Achievement Award.
\end{IEEEbiography}

\begin{IEEEbiography}
[{\includegraphics[width=1in,height=1.25in,clip,keepaspectratio]{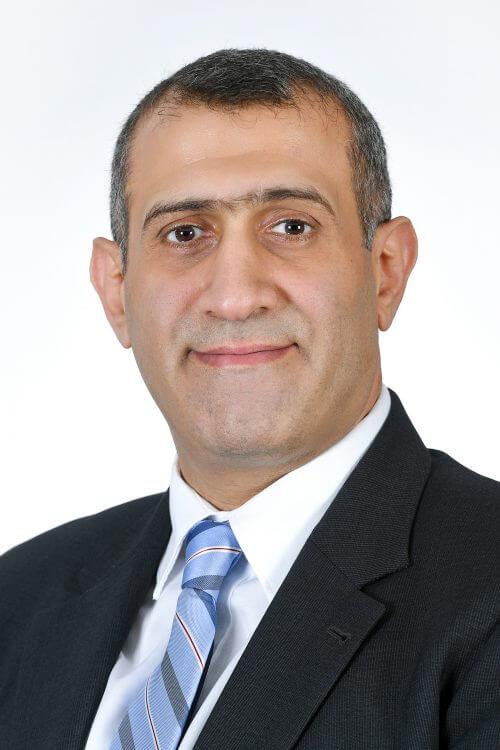}}]%
{Hadi Otrok}  (senior member, IEEE) received his Ph.D. in ECE from Concordia University. He holds a Full professor position in the department of Electrical Engineering and Computer Science (EECS) at Khalifa University, an affiliate associate professor in the Concordia Institute for Information Systems Engineering at Concordia University, Montreal, Canada, and an affiliate associate professor in the electrical department at Ecole de Technologie Superieure (ETS), Montreal, Canada. He is an associate editor at: IEEE TNSM, Ad-hoc networks (Elsevier), and IEEE TSC. His research interests include: Blockchain, reinforcement learning, Federated Learning, crowd sensing and sourcing, ad hoc networks, and cloud and fog security.
\end{IEEEbiography}

\begin{IEEEbiography}
[{\includegraphics[width=1in,height=1.25in,clip,keepaspectratio]{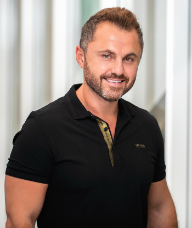}}]%
{Azzam Mourad} Received his M.Sc. in CS from Laval University, Canada (2003) and Ph.D. in ECE from Concordia University, Canada (2008). He is currently Professor of Computer Science and Founding Director of the Cyber Security Systems and Applied AI Research Center with the Lebanese American University, Visiting Professor of Computer Science with New York University Abu Dhabi and Affiliate Professor with the Software Engineering and IT Department, Ecole de Technologie Superieure (ETS), Montreal, Canada. His research interests include Cyber Security, Federated Machine Learning, Network and Service Optimization and Management targeting IoT and IoV, Cloud/Fog/Edge Computing, and Vehicular and Mobile Networks. He has served/serves as an associate editor for IEEE Transactions on Services Computing, IEEE Transactions on Network and Service Management, IEEE Network, IEEE Open Journal of the Communications Society, IET Quantum Communication, and IEEE Communications Letters, the General Chair of IWCMC2020-2022, the General Co-Chair of WiMob2016, and the Track Chair, a TPC member, and a reviewer for several prestigious journals and conferences. He is an IEEE senior member.
\end{IEEEbiography}

\vfill

\end{document}